\documentclass{aa}

\usepackage{graphicx}
\usepackage{xcolor}
\usepackage{txfonts}
\usepackage{orcidlink}
\usepackage{soul}
\begin{document}

   \title{B-FROST: B-Fields and Star Formation across Scales with TRAO}

  \subtitle{CO Abundances, Dynamics and Relative Orientations in the Translucent High Latitude Cloud MBM12}

   \author{J.M. Vorster\inst{1}\orcidlink{0000-0001-9572-2425} 
          \and
          J. Montillaud\inst{2} 
          \and
          M. Juvela\inst{1} 
          \and 
          E. Falgarone\inst{18}
          \and
          J. Oers\inst{3}  
          \and
          E. Mannfors\inst{1} 
          \and
          D. Alina\inst{4} 
          \and        
          Q. Gu\inst{5,6} 
          \and
          H. Kang\inst{7,8} 
          \and
          C.W. Lee\inst{7,9}  
          \and
          S. Li\inst{5,6} 
          \and
          T. Liu\inst{10} 
          \and
          K. Pattle\inst{11} 
          \and
          V.-M. Pelkonen\inst{12,13}\orcidlink{0000-0002-8898-1047} 
          \and
          I. Ristorcelli\inst{3} 
          \and
          A. Zavagno\inst{15,16}
          \and
          L. V. Tóth\inst{14,17}  
          }

        \institute{Department of Physics, University of Helsinki, Finland\\
              \email{jakobus.vorster@helsinki.fi}
              \and
              {Institut UTINAM – UMR 6213 – CNRS – Univ. Bourgogne Franche Comté, OSU THETA, 41bis avenue de l’Observatoire, 25000 Besançon, France}
              \and
              {IRAP, Universit{\'e} de Toulouse, CNRS, Toulouse, France
              }
              \and
              {Physics Department, School of Sciences and Humanities, Nazarbayev University, Astana, Kazakhstan.
              }
              \and
              {School of Astronomy and Space Science, Nanjing University, Nanjing, People’s Republic of China
              }
              \and
              {Key Laboratory of Modern Astronomy and Astrophysics (Nanjing), Ministry of Education, Nanjing, People’s Republic of China
              }
              \and
              {Korea Astronomy and Space Science Institute, Daejeon, Republic of Korea
              }
              \and
              {Astronomy Program, Department of Physics and Astronomy, Seoul National University, Seoul, Korea
              }
              \and
              {University of Science and Technology, Korea (UST), Daejeon, Republic of Korea
              }
              \and
              {Shanghai Astronomical Observatory, Chinese Academy of Sciences, Shanghai, People’s Republic of China
              }
              \and
              {Department of Physics and Astronomy, University College London, London, United Kingdom
              }
              \and
              {Institut de Ciències del Cosmos, Universitat de Barcelona, Barcelona, Spain.
              }
              \and
              {INAF - IAPS, via Fosso del Cavaliere, 100, Roma, Italy
              }
              \and
              {Institute of Physics and Astronomy, ELTE Eötvös Loránd University, Budapest, Hungary
              }
              \and
              {Aix Marseille Univ, CNRS, CNES, LAM, Marseille, France
              }
              \and
              {Institut Universitaire de France, 1 rue Descartes, Paris
              }
              \and
              {Institute of Physics, University of Debrecen, Hungary}
              \and
              {LPENS, Ecole Normale Supérieure, Université PSL, CNRS, Sorbonne Université, Université de Paris, France}
             }

   \date{Received February 26, 2026; accepted May 18, 2026}

 
  \abstract
   {On average, in our Galaxy, the star formation efficiency (SFE) is of the order of a few percent, lower than theoretical predictions. Detailed observational studies of individual molecular clouds may highlight the contributing factors to the low galactic SFE.}
   {We investigated the high-latitude molecular cloud MBM12 as part of the B-fields and star formation across scales (B-FROST) survey with the Taeduk Radio Astronomical Observatory (TRAO) to assess why star formation activity in MBM12 is  low.}
   {We estimated $N$(H$_2$) with {\it Herschel} dust emission and a dust opacity $\kappa_{\nu}$ derived from near-infrared extinction, 21 cm HI column densities and far-infrared emission. With $^{12}$CO and $^{13}$CO ($J=1-0$) line observations, covering an area of $2.5^\circ \times 3^\circ$  at $48''$ resolution we mapped the CO column density $N$(CO), CO-to-H$_2$ factor $X$(CO), and abundance [CO/H$_2$]. We estimated the multi-scale virial parameter $\alpha_{\rm vir}$ and constructed mass-size scaling laws of hierarchical structures with dendrograms. We computed the relative orientation between column density structures and magnetic fields using {\it Planck} observations of dust polarisation.}
   {We identify four main regions based on velocities with H$_2$ column densities ranging from $2\times10^{20}$ cm$^{-2} - 1.3\times10^{22}$ cm$^{-2}$. The CO integrated line intensity, $W$(CO), increases linearly with $N$(H$_2$) providing an average $X$(CO) factor close to the galactic average. At low $N$(H$_2$), $X$(CO) varies below $X_{\rm Gal}$ due to the fall-off of collisional de-excitation in low density gas, and above $X_{\rm Gal}$ due the drop of CO abundances in poorly shielded cloud edges. The hierarchical structures follow a broken power law mass-size relation $M=AR^\alpha$. The values of $\alpha_{\rm vir}$ ranged from $3-60$, with the smallest values  at 0.1 pc scales. The mass-size relations for the structures with the lowest $\alpha_{\rm vir}$ have scaling factors $A$ three times larger than those of high $\alpha_{\rm vir}$ structures, indicating external pressure one order of magnitude larger than the former. We found a transition of parallel to perpendicular between column density structures and magnetic field orientations at $N$(H$_2$) $= 4.5 \times 10^{21}$ cm$^{-2}$.}
   {We provide the first integrated chemical, dynamical, and magnetic field analysis of MBM12. Further investigation into the scale dependence of the mass-size relation and virial parameter can highlight the role of external pressure in regulating the star-formation efficiency.}

   \maketitle
   \nolinenumbers
   
%
    
\section{Introduction}
\begin{figure*}
    \centering
    \includegraphics[width=\linewidth]{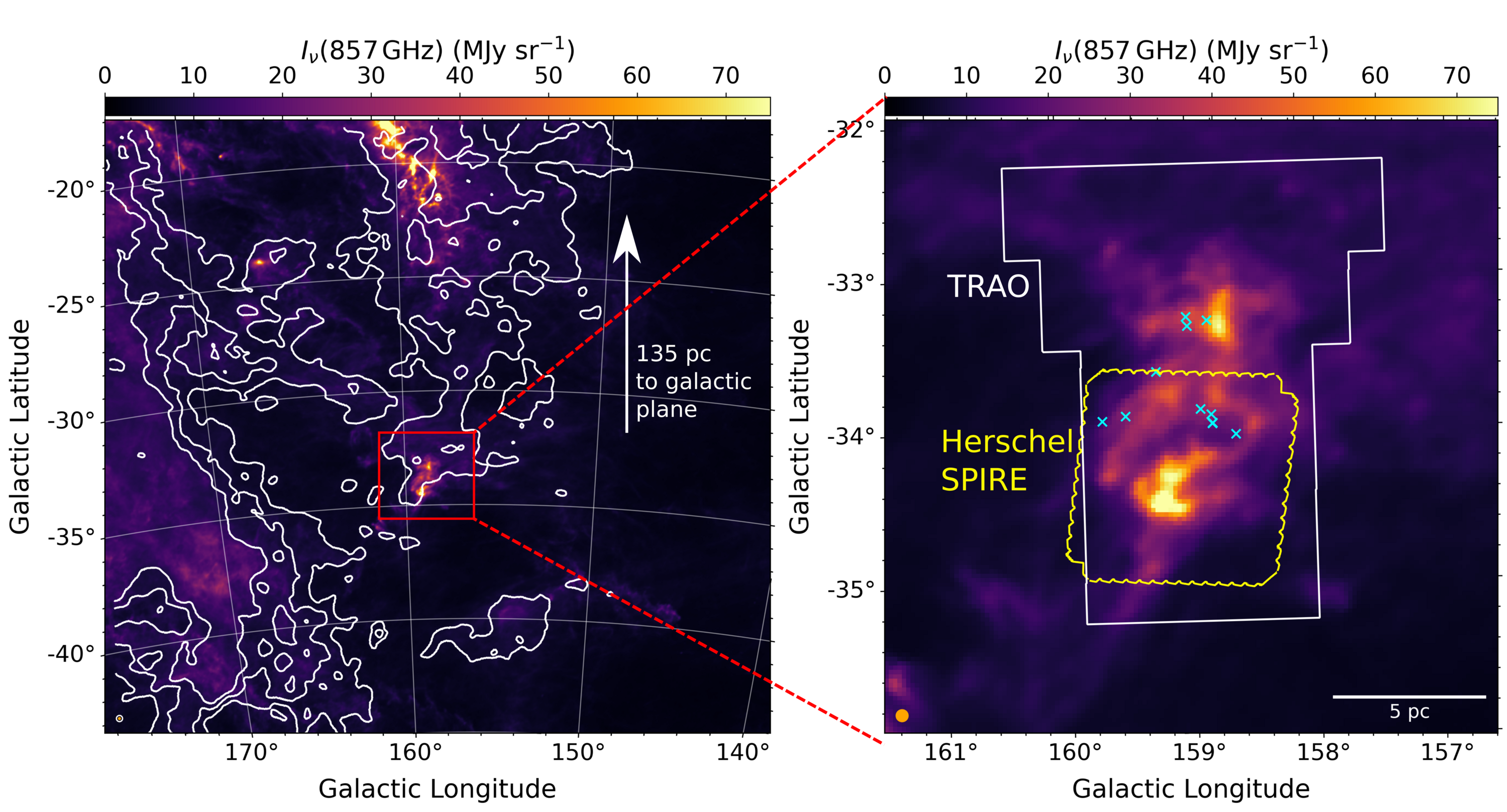}
    \caption{Left: Position of MBM12 relative to the galactic disk. The colour scale shows {\it Planck} 857 GHz dust continuum and the contours are neutral hydrogen column density with contours at levels [1.0, 1.2, 1.4] $\times 10^{21}$ cm$^{-2}$ from the HI4PI all sky survey \citep{2016A&A...594A.116H}. The beam sizes of {\it Planck} 857 GHz (orange) and H14PI (white) are shown in the bottom left. Assuming a distance of 252 pc to MBM12 and a solar height of 20 pc above the galactic plane, the distance of MBM12 from the galactic plane is shown in white. Right: Zoom-in of MBM12 with the positions of T Tauri stars from \citet{2009A&A...497..379M} shown with cyan [$\times$] markers. The observing field of view of the TRAO and {\it Herschel} observations used in this work are shown in white and yellow respectively. A distance scale is shown in the bottom right.}
    \label{fig:mbm12_large_scale}
\end{figure*}
Molecular clouds (MCs) are among the densest and coldest large-scale structures in the interstellar medium (ISM) and the sites of star formation in galaxies. They evolve under the combined effects of self-gravity, turbulence, magnetic fields, radiation and stellar feedback \citep{2012A&ARv..20...55H}. The chemical evolution of MCs is closely linked to their dynamics due to comparable dynamical and chemical timescales   \citep{2014prpl.conf....3D,2023ASPC..534....1C}. The most commonly used tracers for large scale ($>1$ pc) molecular cloud properties are carbon monoxide (CO) line emission, near-infrared (NIR) extinction, and far-infrared (FIR) dust emission, each offering complementary insights. 
\noindent Galactic surveys of $^{12}$CO ($J = 1-0$) line emission have been used for over forty years to estimate MC masses, sizes, surface densities, densities and velocity dispersions. These measurements underpin many of the classic MC scaling relations \citep{1970ApJ...161L..43W,1987ApJ...319..730S,2001ApJ...547..792D,2010ApJ...723..492R,2017ApJ...834...57M}. 

Near-infrared extinction is another dust-based tracer of MC structure, though the consequent column densities depends on the assumed extinction law, but is independent of dust temperature variations along the line of sight\citep{1994ApJ...429..694L,2001Natur.409..159A,2018A&A...620A..24H}. With the all-sky 2MASS survey, it is possible to produce $\sim 5'$ resolution extinction maps for most of the sky \citep{1997ASSL..210...25S,2001A&A...377.1023L,2009A&A...493..735L,2016A&A...585A..38J}. The achievable resolution and dynamic range of column densities from NIR extinction are limited by stellar density. 

Far-infrared dust emission is not limited by background stellar density. The {\it Planck} satellite observed the entire sky at $5'$ resolution \citep[e.g.][]{2014A&A...571A..11P}, while the {\it Herschel} satellite allowed sub-arcmin resolution and two-order-of-magnitude range in column density \citep{2010A&A...518L..93J,2010A&A...518L.102A,2011A&A...527A.111J}. Polarized dust emission also enables measurements of magnetic field orientation \citep{2023ASPC..534..193P}. 

Each tracer presents unique advantages and limitations. CO line emission can probe gas dynamics and chemistry, but transitions between sub-thermal, thermal, optically thick, and self-absorbing regimes complicate mass and column density estimates \citep[e.g.][]{2009ApJ...692...91G}. Extinction is among the most reliable tracers of hydrogen column density $N$(H) for a constant dust to gas ratio, though the extinction law $R_V$ may vary and the attainable resolution is limited. FIR dust emission offers high resolution and dynamic range but depends on variable dust opacities and requires high-altitude or space-based observations. Magnetic field measurements are constrained by sensitivity and resolution, but they provide critical insight into the magnetohydrodynamic (MHD) processes shaping MCs \citep{2023ASPC..534..193P}. A key limitation of all dust-based tracers is their inability to resolve structures along the line of sight at high resolution \citep[recent 3D extinction maps are addressing this problem at scales $>$ 10 pc,][]{2023ASPC..534...43Z}. 

With the advent of wide-field surveys and improved resolution, it is now possible to combine multiple tracers across large MC samples to test abundance variations, excitation conditions, and structural properties \citep{2006A&A...454..781L,2009ApJ...692...91G,2010ApJ...721..686P,2015ApJ...805...58K,2017A&A...599A..98P,2021ApJ...908...76L}. However, such multi-tracer analyses are rarely conducted systematically across large samples, with a few exceptions \citep[e.g.][]{2022ApJ...931....9L,2023A&A...679A...4S}. The B-Fields and staR fOrmation across Scales with TRAO (B-FROST) survey has completed over 2500 hours of $^{12}$CO and $^{13}$CO ($J=1-0$) observations, covering nine star-forming regions across 90 square degrees (Montillaud et al., in prep). B-FROST aims to characterize the relationship between CO chemistry, multi-scale dynamics, and magnetic fields across a diverse set of environments: from high-mass star-forming regions (Monoceros OB1, W40), and feedback-dominated regions (two fields in the $\lambda$ Orionis ring), to low-mass isolated clouds (MBM12, L183), and compact globules (G110-13, L1780). These sources are at distances of 100 to 700 pc, corresponding to spatial resolutions from $\sim$0.03 pc to $\sim$0.15 pc respectively. 

In single-tracer surveys, environmental diversity can be a weakness, introducing unknown systematics. However, because B-FROST explicitly seeks to understand how CO chemistry and dynamics interact with magnetic fields across environments, this diversity becomes a strength. Therefore, conducting in-depth studies across diverse environments allows us to identify both general trends in star-forming regions and environment-specific variations.

The most basic CO observable is the line area $W$(CO) = $\sum_{\rm chan} T_{\rm b}\Delta v$, with $T_{\rm b}$ the brightness temperature and $\Delta v$ the channel width. The line area is related to the molecular hydrogen column density $N$(H$_2$) through $X$(CO) $= N$(H$_2$)/$W$(CO). For $^{12}$CO this factor has a Galactic average value of $X_{\rm Gal} = 2\times10^{20}$ cm$^{-2}$ (K km s$^{-1}$)$^{-1}$ \citep{2013ARA&A..51..207B}. This conversion factor is essential in extragalactic studies, where high resolution CO observations are often the only way to trace mass \citep[e.g.][]{2018ApJ...860..172S,2024ARA&A..62..369S,2025ApJ...985...14L}. The $X$(CO) factor is normally calculated as a means to an end (to convert CO intensity to mass), but it is rarely used as a tracer of physical properties in and of itself. It is well known that it varies on intra-cloud scales \citep[e.g.][]{2015ApJ...805...58K,2022ApJ...931....9L}. By definition, it is the H$_2$ column density per CO photon, and is therefore connected to processes of H$_2$ and CO abundance variation, as well as CO line excitation and radiative transfer. 

The CO column density $N$(CO) is another fundamental measure. With various assumptions, one can estimate the total CO content of a molecular cloud (see Sect. \ref{subsec:nco_methods}). When combined with independent measures of $N$(H$_2$), one can get the projected CO abundance [CO/H$_2$]. The CO abundance has been used to detect CO freezing onto dust grains in the coldest and densest parts of molecular clouds \citep{2010ApJ...721..686P,2021ApJ...908...76L}. Although CO chemistry has been studied for decades, the relative role of UV-driven CO chemistry with dynamically induced chemistry is still an open question \citep{1988ApJ...334..771V,2012A&ARv..20...55H}. Photodissociation region models of turbulent molecular clouds fail to reproduce observed abundances in translucent gas  \citep{2012A&A...544A..22L}. Further, high resolution maps of abundance have not yet been utilized on a large scale.

Complementary to environmental characteristics of MCs are their dynamics. Self-gravity competes with internal turbulence, magnetic tension, magnetic pressure gradients and stellar feedback \citep{2023ASPC..534....1C}. Molecular clouds are characterized by self-gravitating filamentary structures \citep[e.g.][]{2010A&A...518L.102A}. There is some disagreement about whether the dominant mechanism for MC structure is self-gravity or supersonic turbulence \citep{2019MNRAS.490.3061V,2020ApJ...900...82P}. There is a number of statistical tests that can be run on a position-position-velocity (PPV) cube, but projection effects and molecular abundances make direct comparison with simulations difficult \citep[e.g. the tools in][]{2019AJ....158....1K}. One controversial but easy to calculate measure of gravitational binding is the virial parameter $\alpha_{\rm vir}$, where super- and sub-virial are relative to $\alpha_{\rm vir} = 2$  \citep{1992ApJ...395..140B}, where
\begin{equation}
    \alpha_{\rm vir} = \frac{5\sigma_{\rm v, st}^2R_{\rm st}}{GM_{\rm st}},
    \label{eq:virial_par}
\end{equation}
with $\sigma_{\rm v, st}$ the 1D velocity dispersion of the structure, $R_{\rm st}$ the size, $G$ the gravitational constant and $M_{\rm st}$ the mass of the structure \citep{1992ApJ...395..140B,2013ApJ...779..185K}. The case of $\alpha_{\rm vir} < 2$ does not necessarily imply "bound" and $\alpha_{\rm vir} > 2$ "unbound" (see the brief discussion in App. \ref{app:virial_analysis_interpretation}). Yet, it can be useful when used in conjunction with other tracers and paired with a physically motivated definition of a cloud. The dendrogram algorithm by \citet{2008ApJ...679.1338R} is a natural way to segment PPV cubes into hierarchical structures. Many analyses utilize the highest levels of dendrogram structures \citep[e.g.][]{2024ApJ...969...70F}, yet only a few utilize the full multi-scale hierarchy produced by the dendrogram algorithm \citep{2025ApJ...993..193O}. Tracing the scale dependence of $\alpha_{\rm vir}$ within a MC structure may yield additional insights into its stability.

Magnetic fields should not be ignored, as they increase in strength with density \citep{2010ApJ...725..466C,2023ASPC..534..193P}. There has been a lot of work done mapping B-fields across scales in molecular clouds \citep[][]{2021A&A...647A..78A,2025ApJ...985..222H}. In many MCs, magnetic fields transition from being parallel to column density structures at low column densities to perpendicular at high column densities, suggesting they influence cloud evolution \citep{2013ApJ...774..128S,2019A&A...629A..96S}. Highly magnetized gas has preferential motion along field lines, leading to a perpendicular arrangement at the strong field, high density limit \citep{2017A&A...603A..64S,2025A&A...698A.119S}. Algorithms such as \texttt{FilDReaMS} allow a multi-scale consideration of B-field to column density relative orientations \citep[][Oers et. al. submitted]{2022A&A...668A..41C,2022A&A...668A..42C}. Magnetic field studies are not often paired with CO virial analyses, but they may complement one another. Virial analysis with CO typically ignores magnetic fields, and relative orientation analysis cannot separate components along the line of sight.

The source we investigate as part of the B-FROST survey in this paper is the high-latitude molecular cloud MBM12 (Fig. \ref{fig:mbm12_large_scale}). MBM12 has an estimated age of $2^{+3}_{-1}$ Myr, and contains about a dozen T Tauri stars \citep{2001ApJ...560..287L,2002AJ....124.3387H,2009A&A...497..379M,2012ApJ...746...11K}. It is at a distance of $252^{+4}_{-6}\pm12$ pc \citep{2019ApJ...879..125Z}, and is a well-studied molecular cloud from the MBM catalogue \citep{1985ApJ...295..402M}. Analysis of large-scale CO mapping assumed a distance of 65 pc, and concluded from a virial analysis that MBM12 is dispersing or pressure-bound \citep{1990ApJ...351..165P}. Analysis of ammonia mapping of MBM12 concluded that it is not forming stars \citep{2000MNRAS.314..743G}. It was also studied as part of the Galactic Cold Cores {\it Herschel} programme, with an estimated mass of 1.2$\times 10^3$ M$_\odot$ \citep{2015A&A...584A..92M}. A follow-up IRAM 30-m survey of one of the cores found rich prestellar chemistry, but did not detect deuterium or oxygen-bearing species, suggesting a quiescent core, or one at the early stages of star formation \citep{2022A&A...658A.131Z}. Virial analysis of a recent CO survey  found $\alpha_{\rm vir} > 30$ for most cores in the cloud \citep{2021ApJ...920..103X}. However, \citet{1997ApJ...475..642M} used CO and HI observations to show that MBM12 might be pressure compressed at its southern end. So although most studies imply that MBM12 has finished forming stars and is dispersing into the environment, it may contain gas at the onset of triggered star formation.

    As part of the B-FROST survey, we present an in-depth case study of MBM12. In this work, we aim to test whether MBM12 shows signs of gravitational instability despite previous indications of dispersal. We also assess how CO chemistry varies across the cloud, and whether magnetic field orientation correlates with structural or dynamical features. By combining $^{12}$CO and $^{13}$CO line emission with {\it Herschel} dust continuum data, we derive spatially resolved estimates of $X$(CO), $N$(CO), [CO/H$_2$] and $\alpha_{\rm vir}$ across 0.05 - 1 pc scales. MBM12’s high-latitude environment allows us to explore the role of magnetic fields and the interplay between CO chemistry and gravitational stability in a relatively pristine setting. In Section \ref{sec:observations} we introduce the data from TRAO, {\it Herschel} and {\it Planck} observations. Then we describe our methods for estimates of H$_2$ column density (Sect. \ref{subsec:nh2}), CO column density (Sect. \ref{subsec:nco_methods}), virial parameters (Sect. \ref{subsec:virial_co_methods}), and histogram of relative orientations (HROs, Sect. \ref{subsec:hro_methods}). In Section \ref{subsec:environmental_variation} we analyze the spatial distribution and probability density functions (PDFs) of $N$(H$_2$), $N$(CO), $X$(CO) and [CO/H$_2$] and compare these quantities against each other. We then look at multi-scale virial parameters (Sect. \ref{subsec:multiscale_dynamics}) and the histogram of relative orientation for MBM12 (Sect. \ref{subsec:results_magnetic_fieldsMBM12}). We discuss our findings in Section \ref{sec:discussion}.

\section{Observations}
\label{sec:observations}
\subsection{TRAO observations}

The B-FROST survey observed MBM12 (l,b)$=(159^\circ00'00''$,$-34^\circ00'00''$) for 288 hours with the TRAO, operated by the Korean Astronomy and Space Science Institute (KASI)\footnote{\url{https://trao.kasi.re.kr/main.php}}. More details on the observations, data reduction and data quality are found in Montillaud et al. (in prep). We observed the $^{12}$CO ($J=1-0$) and the $^{13}$CO ($J=1-0$) molecular lines simultaneously in on-the-fly mode with a spectral resolution of 0.2 km s$^{-1}$. The angular resolution was between 46$''$ and 48$''$ for 110 GHz $-$ 115 GHz, and pointing accuracy was better than 10$''$\citep{2019JKAS...52..227J}. We resampled all data to a grid resolution of 44$''$. MBM12 was mapped by scanning $0.6^\circ \times 0.6^\circ$ tiles in orthogonal $l$ and $b$ directions. The average system temperature was 300 K and 660 K at 110 GHz and 115 GHz respectively, but we repeated the scans until the desired noise level was reached. The final map size was $2.5^\circ\times3^\circ$. The tiles were gridded and averaged with the GILDAS CLASS\footnote{\url{https://www.iram.fr/IRAMFR/GILDAS}} software to make a single spectral cube for each line. We achieved a $T_{\rm A}^*$ sensitivity of $\sigma_{\rm ^{12}CO} = 0.28$ K and $\sigma_{\rm ^{13}CO} = 0.13$ K for most of the tiles, although some areas had a higher noise level. We converted the antenna temperature $T_{\rm A}^*$ to main beam temperature $T_{\rm mb}$ by dividing the antenna temperature by $\eta_{\rm eff} = 0.51$ at 115 GHz and $\eta_{\rm eff} = 0.54$ at 110 GHz. 
\subsection{Herschel and Planck observations}
MBM12 was observed with the {\it Herschel} Space Observatory Spectral and Photometric Imaging REceiver \citep[SPIRE,][]{2010A&A...518L...3G} at 250 $\mu$m, 350 $\mu$m and 500 $\mu$m as part of the Galactic Cold Cores open time programme \citep{2010A&A...518L..93J}. We retrieved the MBM12 maps from the {\it Herschel} archive. The resolutions for these maps are about 18$''$, 25$''$, and 37$''$. The relative calibration accuracies of the {\it Herschel} SPIRE surface brightness maps are expected to be better than 4$\%$ \citep{2013MNRAS.433.3062B}\footnote{\url{https://www.cosmos.esa.int/web/herschel/ao2-documentation}}.
We also used the \textit{Planck} 353 GHz full mission data, which are available as part of the Product Release 2018 in the \textit{Planck} Legacy Archive \footnote{\textit{Planck} Legacy Archive: \url{https://www.cosmos.esa.int/web/planck/pla}.}, observed with the \textit{Planck} High Frequency Instrument (HFI, \citealp{Lamarre2010}). We extracted from the all-sky \textit{Planck} data 2$^{\circ} \times 2^{\circ}$ maps of the Stokes total intensity \textit{I} and the \textit{Q} and \textit{U} parameters, which correspond to the linear polarisation components, and their uncertainties at 353 GHz centred on the \textit{Herschel} maps of MBM12. We smoothed the \textit{Planck} maps from an angular resolution of $4.7^{\prime}$ to $7^{\prime}$ to improve the signal-to-noise ratio. Appendix \ref{app:polarization_angle} describes our estimates of the polarisation angles.
\section{Methods}
\label{sec:methods}
\subsection{H$_2$ column density from dust emission}\label{subsec:nh2}
\begin{figure}
    \centering
    \includegraphics[width=0.8\linewidth]{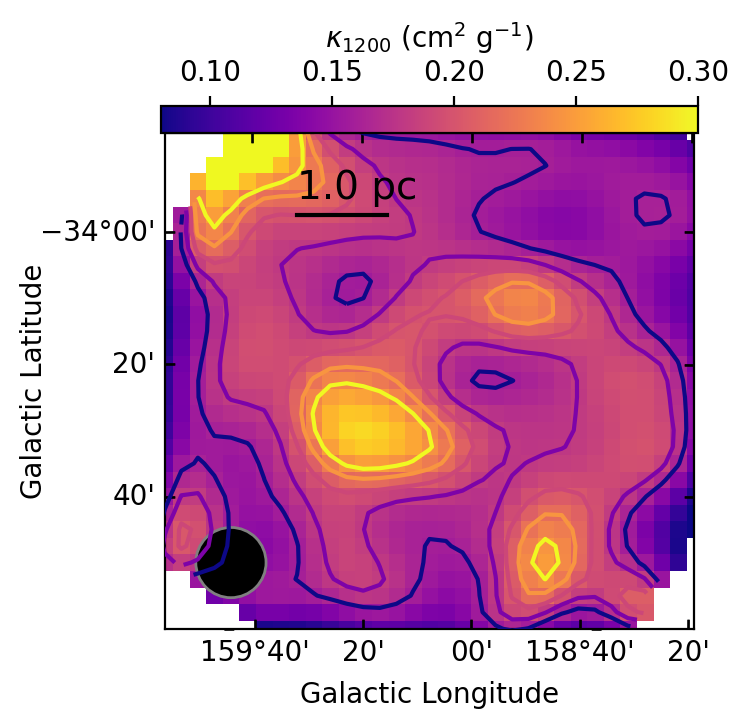}
    \caption{Spatial variation in the 250 $\mu$m dust opacity in MBM12, $\kappa_{\rm 1200}$, derived from the ratio of $\tau_{\rm 1200}$ and $A_{\rm K}$, accounting for variable molecular gas fraction. The conversion assumes an extinction curve with $R_{\rm V} = 3.1$, and $N$(H)/$A_{\rm K} = 1.67\times10^{22}$ cm$^{-2}$ mag$^{-1}$ \citep{1989ApJ...345..245C,1978ApJ...224..132B,2022ApJ...931....9L}. Contours of $\kappa_{\rm 1200} = [0.16, 0.18, 0.2,0.22,0.24]$ cm$^2$ g$^{-1}$ are overplotted.}
    \label{fig:kappa_spatial_variation}
\end{figure}

The {\it Herschel} SPIRE $250\,\mu$m, $350\,\mu$m and $500\,\mu$m maps were fit with modified black body (MBB) functions. Maps were colour corrected, background subtracted and convolved to an angular resolution of $40''$. The background subtraction was done by subtracting the average surface brightness of low-emission region centred on $(\alpha,\delta) = (2^{\rm h}53^{\rm m}56^{\rm s}, +19^\circ19'17'')$ with a radius of $4'$ from each map. If the dust emission is optically thin the intensity $I_\nu$ can be written as:
\begin{equation}
    I_\nu \approx B_\nu(T_\text{dust})\tau_\nu,
\end{equation}
with $B_\nu(T_{\rm dust})$ a black body intensity with dust temperature $T_{\rm dust}$. The values of $\tau_\nu$ and $T_{\rm dust}$ for each pixel, as well as their uncertainties, were estimated with Markov Chain Monte Carlo (MCMC) runs as in \citet{2015A&A...584A..93J}. The H$_2$ column density, $N$(H$_2$), can then be estimated from:
\begin{equation}
    \label{eq:tauNH2}
    N{\rm (H}_2{\rm )} = \frac{\tau_\nu}{\kappa_\nu \mu_{\rm H_2}m_{\rm p}},
\end{equation}
where $\tau_{\rm \nu}$ is the dust optical depth at a frequency $\nu$, $\kappa_\nu$ the dust opacity in cm$^2$ g$^{-1}$ assuming a dust-to-gas ratio of 100, $m_{\rm p}$ the proton mass, and the molecular mass per H$_2$ molecule $\mu_{\rm H_2} = 2.74$ \citep[e.g.][]{2010ApJ...723.1019H}. The opacity is known to vary by a factor of three between the dense and diffuse ISM \citep{2014A&A...571A..11P,2015A&A...577A.110Y,2015A&A...584A..93J}. Our source, MBM12, is a high-latitude cloud, so assuming a literature value of $\kappa_0$ for the partly translucent ISM may be unreliable Therefore, we explored empirical calibration of $\kappa_\nu$. Calibration of $\kappa_\nu$ to use with FIR emission requires a reference column density estimate from an independent tracer. We calibrated the 250 $\mu$m (1200 GHz) optical depth $\tau_{\rm 1200}$ against the K-band extinction $A_{\rm K}$ \citep{2014A&A...566A..45L,2022ApJ...931....9L}. For a \citet{1989ApJ...345..245C} extinction curve, $R_{\rm V} = 3.1$, and an extinction to atomic hydrogen column density conversion factor $\beta_{\rm K} \equiv N({\rm H})/A_{\rm K} = 1.67\times10^{22}$ cm$^{-2}$ mag$^{-1}$, the dust opacity for $N$(H$_2$) at 1200 GHz is given by \citep{2022ApJ...931....9L}
\begin{equation}
    \kappa_{\rm 1200} = \frac{1}{f_{\rm mol}\mu_{\rm H_2}m_{\rm p}\beta_{\rm K}\gamma_{\rm 1200}}
    \label{eq:extinction_kappa}
\end{equation}
where $\gamma_{\rm 1200} = A_{\rm K}/\tau_{\rm 1200}$ and $f_{\rm mol}$ is the molecular gas fraction. We extracted a K-band extinction map from the iNICEST online platform \citep{2001A&A...377.1023L,2009A&A...493..735L}\footnote{\url{http://interstellarclouds.fisica.unimi.it/html/index.html}}. We chose a control field close to MBM12, and a smoothing FWHM of 5$'$, and a pixel size of 2.5$'$. We then convolved and reprojected the $\tau_{\rm 1200}$ map to the same grid and used Eq. \ref{eq:extinction_kappa} to produce the spatial variation in $\kappa_{\rm 1200}$ (Fig. \ref{fig:kappa_spatial_variation}). We also took spatial variations in $f_{\rm mol}$ into account (App. \ref{app:molecular_gas_fraction}). This $N$(H$_2$) is then the molecular content, rather than the total hydrogen content along the line of sight. We are only considering the molecular cloud MBM12, as CO emission is emitted from inside a molecular cloud.

\subsection{LTE CO column density}
\label{subsec:nco_methods}
The method used to derive CO column densities for MBM12 is given in Montillaud et al. (in prep). We give a brief overview of the methodology here. We derived the CO column densities, $N$(CO), with large-scale TRAO CO line observations. We followed \citet{2010ApJ...721..686P} to estimate $N$($^{13}$CO). The main steps of the LTE estimate were estimating a $T_{\rm ex}$ map from the peak $T_{\rm mb}$ of each pixel's spectrum in the $^{12}$CO map, assuming the same $T_{\rm ex}$ between $^{12}$CO and $^{13}$CO \citep[Eq. 19 of][]{2010ApJ...721..686P}. After the $T_{\rm ex}$ estimate, the $^{13}$CO optical depth $\tau^{13}(v)$ was calculated for each channel width $T_{\rm mb}$ and the $T_{\rm ex}$\citep[Eq. 20 of][]{2010ApJ...721..686P}. Lastly, $^{13}$CO column density was calculated from $\tau^{13}(v)$ and $T_{\rm ex}$ \citep[Eq. 17 of][]{2010ApJ...721..686P}. Then $N$($^{13}$CO) was converted to $N$(CO) by using the empirical isotopologue ratio with Eq. 4 of \citet{2014MNRAS.445.4055S}, with the coefficients of model e in Table 3 of that paper. The different models consist of different initial conditions in their simulations. The choice of the model for the isotope ratio has around 20$\%$ systematic uncertainty, but it is much smaller than the uncertainty from assuming a constant isotope ratio \citep{2014MNRAS.445.4055S}. The assumption of the same excitation temperature between $^{12}$CO and $^{13}$CO adds uncertainty to our estimates \citep{2000ApJ...529..259P}. To improve this estimate, one could either model the conversion of $T_{\rm ex}$ between two isotopologues, or one could survey the cloud in the $^{13}$CO ($J=2-1$) line. However, modelling the difference of $T_{\rm ex}$ between the two isotopologues is non-trivial, and we did not have access to the $^{13}$CO ($J=2-1$) line at $2.5^\circ \times 3^\circ$ scales.
\subsection{Virial analysis with CO}
\label{subsec:virial_co_methods}
\subsubsection{Dendrogram generation}
We applied the dendrogram algorithm \citep{2008ApJ...679.1338R} with the \texttt{astrodendro} package to both CO spectral cubes. The algorithm assigns each voxel in the position-position-velocity (PPV) cube to one or more structures of connected isocontours. The structures are nested, with higher $T_{\rm mb}$ structures in PPV space always within lower $T_{\rm mb}$ structures. The trunk contains all the structures, while branches are structures with more structures within them. Structures that contain no further hierarchical structures are leaves. The minimum antenna temperature for a voxel to be considered for a structure is $T_{\rm min}$. The minimum number of voxels for a structure is determined by the parameter $N_{\rm min}$. The final parameter, $\Delta T_{\rm min}$, is the minimum $T_{\rm mb}$ difference between two peaks in PPV which could be considered as separate structures. We took $N_{\rm min} = 27$, $T_{\rm min} = 3\sigma_{\rm rms}$, and $\Delta T_{\rm min} = 3\sigma_{\rm rms}$, following the suggestion of \citet{2008ApJ...679.1338R}. 
\subsubsection{Virial parameter estimates}
\label{subsubsec:methods_virial_par}
We estimated the virial parameter $\alpha_{\rm vir}$  for dendrogram structures assuming each structure corresponds to a sphere with mass $M_{\rm st}$ and radius $R_{\rm st}$. We assume optically thin $^{13}$CO emission, corresponding to a direct conversion factor between the sum of $T_{\rm mb}$ of the structure and the mass $M_{\rm st}$ (Eq. \ref{eq:cloud_mass}). The virial parameter $\alpha_{\rm vir}$ was calculated for structures at all levels with Eq. \ref{eq:virial_par}. We used a locally calibrated $\langle X$($^{13}$CO)$\rangle = 7.83\times10^{20}$ cm$^{-2}$ (K km s$^{-1}$)$^{-1}$ to convert $T_{\rm mb}$ to $N$(H$_2$) (estimated in Sect. \ref{subsec:environmental_variation} and Fig. \ref{fig:environment_summary}f). The benefit of using line observations with an empirical conversion from $T_{\rm mb}$ to $N$(H) to estimate $\alpha_{\rm vir}$ is that velocity components can be considered individually, under the assumption that each velocity component corresponds to a respective gas clump. Separating structures into a dendrogram avoids spurious increases in $\sigma_{\rm v, st}$ due to multiple velocity components along the line of sight. The mass estimate from the spectral cube was calculated as \citep{2020A&A...638A..74L}
\begin{equation}
    M_{\rm st} = \delta A_{\rm pixel} \mu_{\rm H_2}m_{\rm H} \langle X({\rm CO}) \rangle  \times \sum_{\rm pixels} \sum_{\rm channels} \Delta v T_{\rm mb},
    \label{eq:cloud_mass}
\end{equation}
with $\delta A_{\rm pixel}$ the area of a single pixel, $m_{\rm H}$ the mass of a hydrogen atom and $\Delta v = 0.2$ km s$^{-1}$ the channel width. The mass of larger hierarchical structure includes the mass of that structure's smaller scales. The pixel area $\delta A_{\rm pixel} = s_{{\rm pixel},l}\,s_{{\rm pixel},b}\,d^2$ with $s_{\rm pixel}$ referring to the pixel scale and $d$ the distance to MBM12. The radius $R_{\rm st}$ was estimated as
\begin{equation}
    R_{\rm st} = \Bigg(\frac{A}{\pi}\Bigg)^{0.5},
\end{equation}
with $A = \sum_{\rm pixels}\delta A_{\rm pixel}$ the total pixel area. The velocity dispersion  per pixel $\sigma_{\rm v,pixel}$ was calculated with
\begin{equation}
    \sigma_{\rm v,pixel} = \sqrt{\frac{\sum_{\rm channels}T_{\rm mb,i}(v_{\rm i} - \bar{v})^2}{\sum_{\rm channels} T_{\rm mb,i}}},
\end{equation}
with $T_i$ the antenna temperature in K, $v_i$ the velocity of a channel $i$, $\bar{v}$ the intensity-weighted mean velocity for the structure. The velocity dispersion $\sigma_{\rm v,st}$ was then the unweighted average of $\sigma_{\rm v,pixel}$ for all sky pixels. 
\noindent The largest systematic uncertainties in our estimate of $\alpha_{\rm vir}$ are spatial variation in $X$(CO), structure morphology, and distance. The $\langle$$X$(CO)$\rangle$ we use is the mean for MBM12; for low-density regions the true $X$(CO) can be significantly higher. The use of a constant $X$(CO) may underestimate the mass in low-density regions. Compared to the flux-weighted size from \citet{2008ApJ...679.1338R}, our estimate gives larger $\alpha_{\rm vir}$ as we use the total area for the radius. These considerations suggest that our estimates of $\alpha_{\rm vir}$ are upper limits \citep[while $\alpha_{\rm vir}$ further neglects external pressure and magnetic fields,][]{2006MNRAS.372..443B}. However, we point out that our use of a calibrated $X$(CO) from $\kappa_\nu$-calibrated $N$(H$_2$) reduces the uncertainty on the observed value of $\alpha_{\rm vir}$ significantly compared to assuming a $X$(CO) value from the literature. The distance we adopt has an uncertainty of 10$\%$, which is systematic across the cloud, and so the uncertainty from distance is likely smaller than that from $X$(CO) and morphology.
\subsection{Histogram of relative orientations}
\label{subsec:hro_methods}
We applied \texttt{FilDReaMS} (Filament Detection and Reconstruction at Multiple Scales; \citealp{2022A&A...668A..41C}) to the {\it Herschel} dust column density maps of MBM12 described in Sect.~\ref{subsec:nh2}. \texttt{FilDReaMS} is designed to detect elongated structures, which we will call filaments hereafter, in an image, and provides information on their widths, their orientations and the robustness of the detection. \texttt{FilDReaMS} uses a model template that has the shape of a rectangular bar (referred to as the model bar) defined by its length $L_{\rm{b}}$, width $W_{\rm{b}}$ and aspect ratio $r_{\rm{b}} = L_{\rm{b}}/W_{\rm{b}}$. We adopt $r_{\rm{b}}$ = 3 \citep{2014MNRAS.444.2507P,2019A&A...621A..42A,2022A&A...668A..42C}, and we consider values of $W_{\rm{b}}$ spanning the range $[(W_{\rm{b}})_{\rm{min}},(W_{\rm{b}})_{\rm{max}}]$, with $(W_{\rm{b}})_{\rm{min}} = 5\,\rm{pix}$ and $(W_{\rm{b}})_{\rm{max}} = 29\,\rm{pix}$ equal to the broadest structure detected in the map. The orientation angle of the model bar, $\psi_{\rm{b}}$, is defined in the range $[-90^{\circ}, +90^{\circ}]$ and follows the IAU convention (Sect.~\ref{app:polarization_angle}).
For each value of $W_{\rm{b}}$, \texttt{FilDReaMS} filters out structures broader than $W_{\rm b}$ and converts the resulting image into a binary map. At each pixel $i$ of the binary map, \texttt{FilDReaMS} retrieves the orientation angle of the model bar that best matches the map, $(\psi_{\rm{b}})_{i}$ and computes the corresponding significance, $S_{i}$, which compares the detected filament to an ideal case (see \citealt{2022A&A...668A..41C} for more details). If $S_{i} > 1$, \texttt{FilDReaMS} confirms the detection of a filament with orientation angle $(\psi_{\rm{f}})_{i} = (\psi_{\rm{b}})_{i}$. Once all the pixels have been treated, \texttt{FilDReaMS} creates a filament mask by multiplying the above binary map with a model bar mask formed by the model bars of all the detected filaments. This filament mask is then applied to the initial image to reconstruct the physical network of detected filaments of bar width $W_{\rm{b}}$ with their true shapes. In case a given pixel belongs to two or more filaments, the filament orientation angle assigned to that pixel is the orientation angle of the most significant filament detected. This process is repeated over the entire range of $W_{\rm{b}}$, leading to filamentary networks of different size scales.
\section{Results}
\begin{figure}
    \centering
    \includegraphics[width=\linewidth]{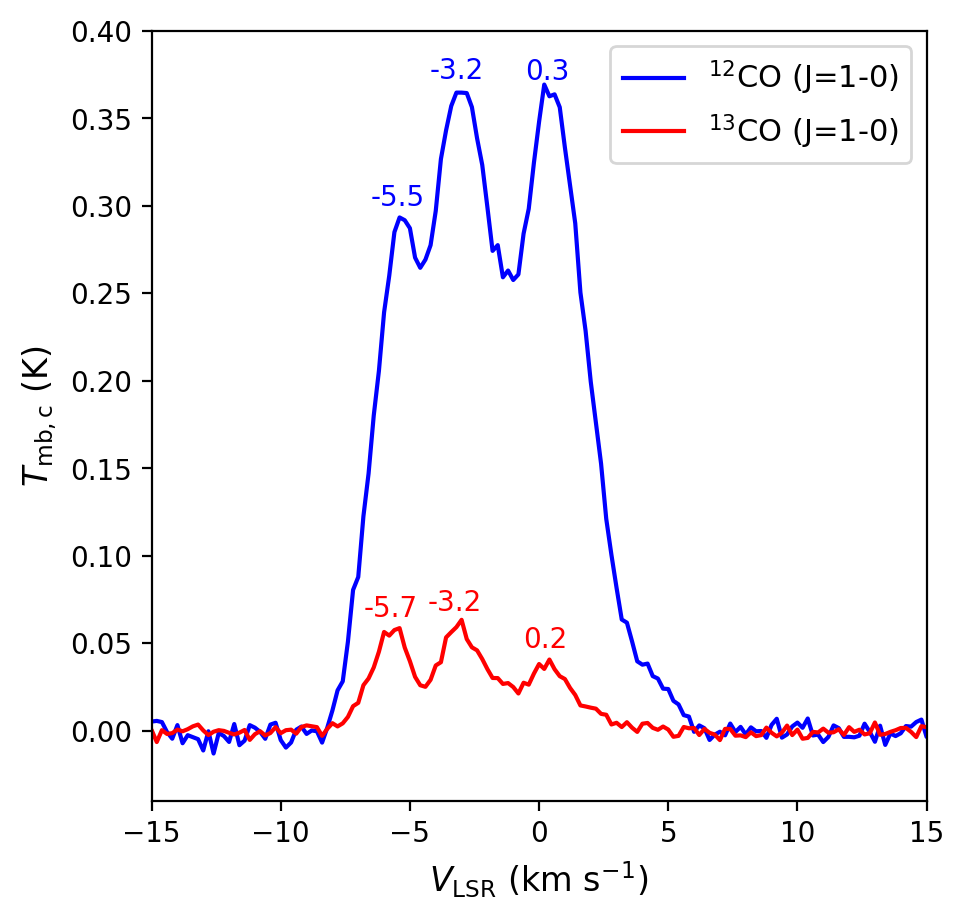}
    \caption{Mean spectra over the entire MBM12 field. The velocities of each peak are indicated.}
    \label{fig:mean-spectrum}
\end{figure}

\begin{figure*}
    \centering
    \includegraphics[width=\linewidth]{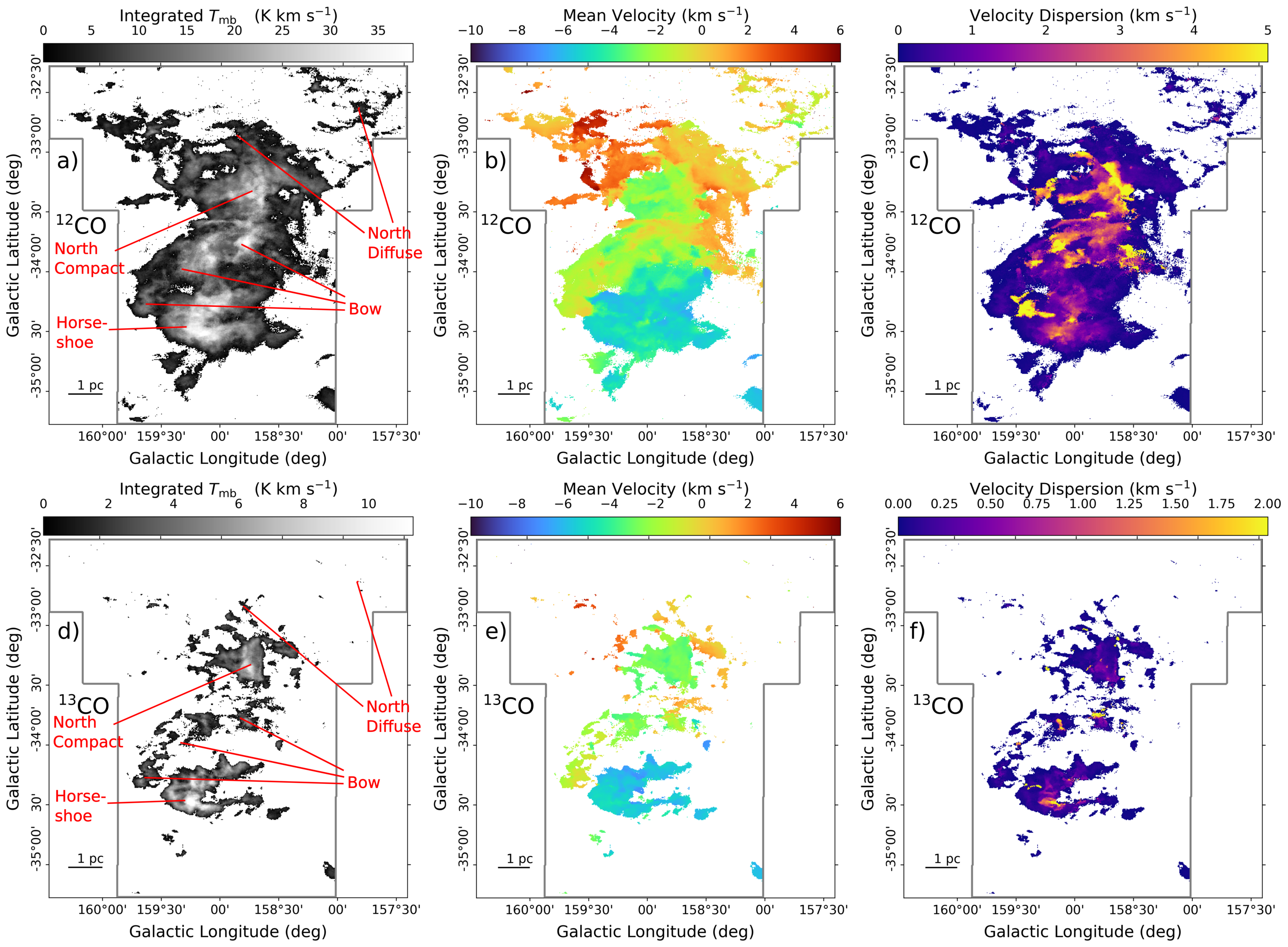}
    \caption{Integrated intensity (a and d), intensity weighted mean velocity (b and e) and intensity weighted velocity dispersion (c and f) for $^{12}$CO (top) and $^{13}$CO (bottom) for MBM12 as observed with the TRAO. The sub-regions of MBM12 referenced in the text are labeled in red.}
    \label{fig:mbm12_momentmaps}
\end{figure*}

We identified four main regions separated by velocities in MBM12: The Horseshoe, Bow, North Compact and North Diffuse (Figs. \ref{fig:mbm12_momentmaps} and \ref{fig:virial_dendro}). Figure \ref{fig:mean-spectrum} shows the mean spectra of the field. These maps were generated with emission $T_{\rm mb} > 5 \sigma_{\rm rms}$. The Horseshoe is an isolated region around $-6$ km s$^{-1}$. The Bow and North Compact are connected in emission in $^{12}$CO but not in $^{13}$CO at around $-2$ km s$^{-1}$. North Diffuse consists of clumpy emission at $> 1$ km s$^{-1}$.

\subsection{Environmental variation}
\label{subsec:environmental_variation}
\begin{figure*}
    \centering
    \includegraphics[width=0.90\linewidth]{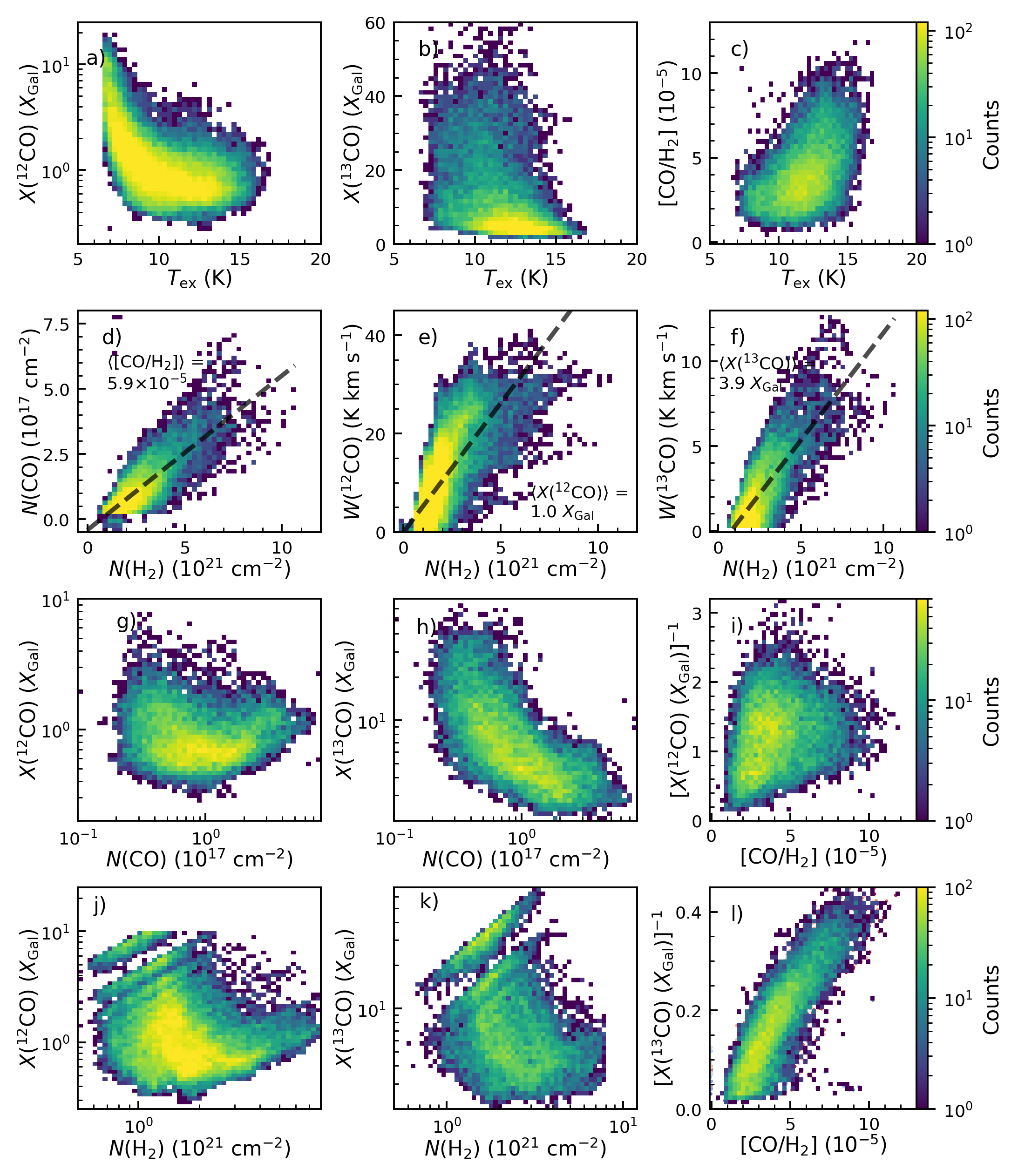}
    \caption{Comparison of H$_2$ and CO quantities observed in MBM12 with Herschel and TRAO. We compare $N$(H$_2$) with CO linewidths $W$(CO), $X$-factors $X$(CO), $^{12}$CO excitation temperature $T_{\rm ex}$, CO column density $N$(CO)$_{\rm lte}$ and CO abundances [CO/H$_2$]$_{\rm lte}$. Panels a)$-$c) and d)$-$l) have different colorscales. Panels d)$-$f) have linear fits shown in black lines, with the slope annotated on the respective panel. The linear striations at low $N$(H$_2$) in panels j) and k) are artefacts of the sigma-clipping in the $W$(CO) estimates.}
    \label{fig:environment_summary}
\end{figure*}

\begin{figure}
    \centering
\includegraphics[width=0.8\linewidth]{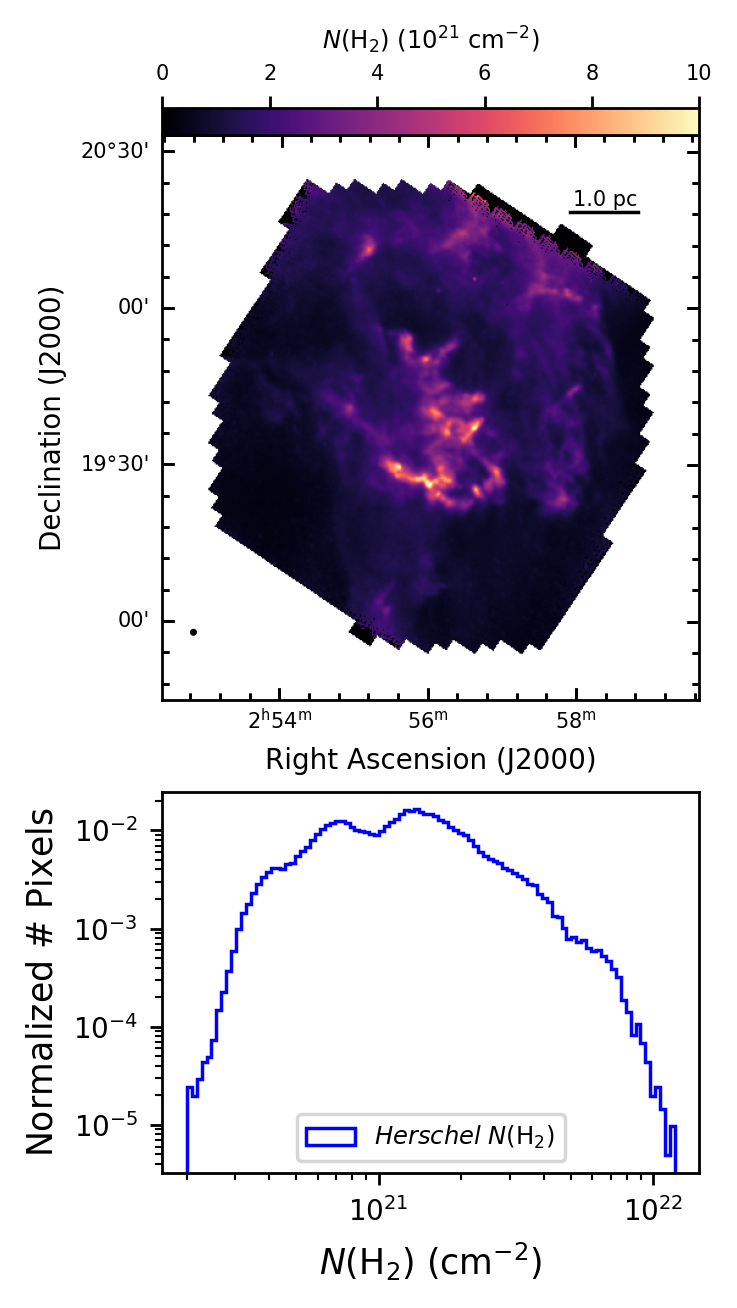}
    \caption{Top: Opacity-calibrated $N$(H$_2$) map from {\it Herschel} observations. Bottom: Dust-opacity-calibrated $N$(H$_2$) PDF with {\it Herschel} for MBM12. The {\it Herschel} map covers only the southern half of MBM12.}
    \label{fig:nh2pdf}
\end{figure}

\begin{figure}
    \centering  \includegraphics[width=\linewidth]{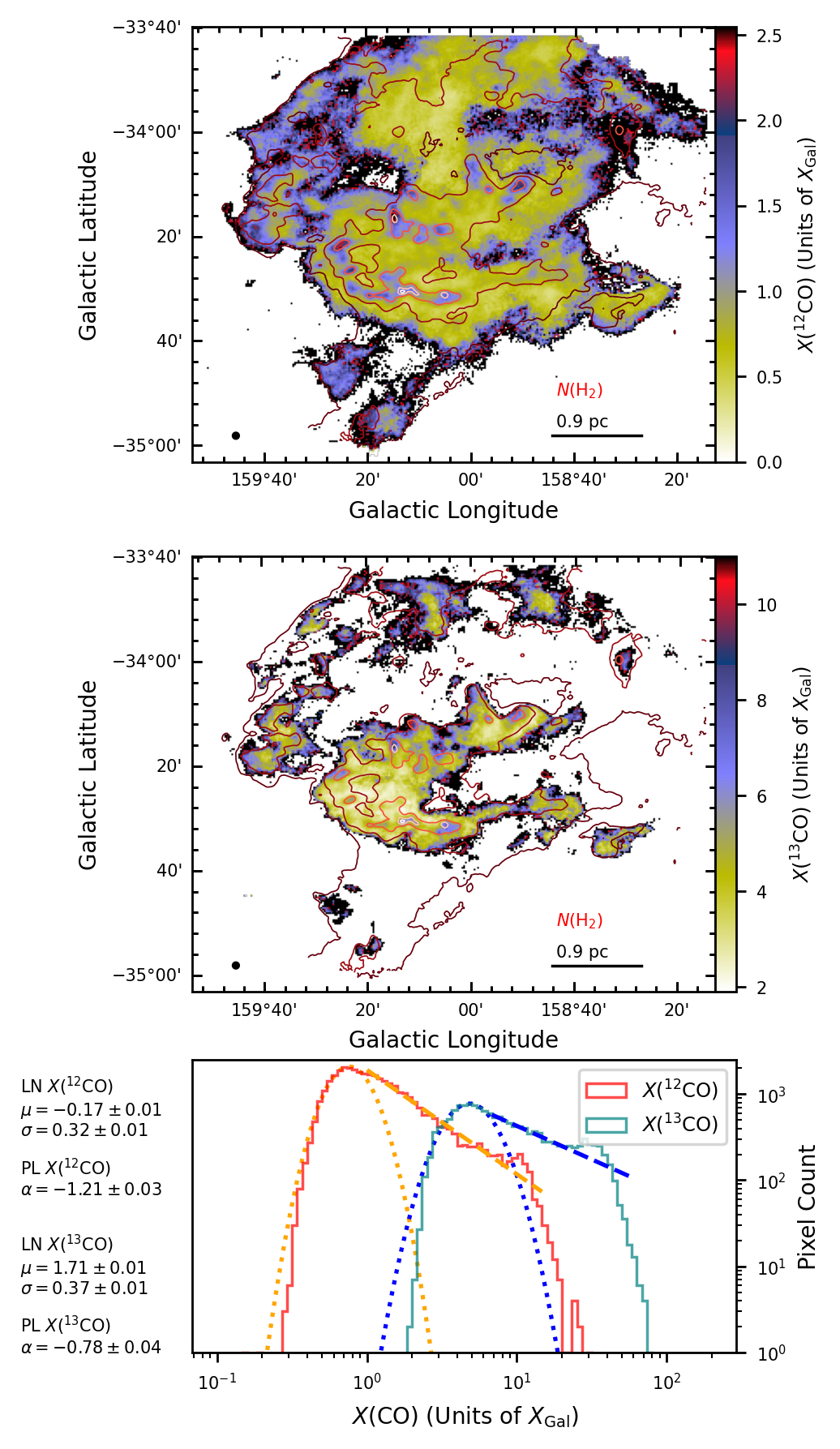}
    \caption{Ratio of $N$(H$_2$) and CO line area for $^{12}$CO and $^{13}$CO ground state rotational transitions in the southern half of MBM12. The figures are in units of $X_{\rm Gal}= 2\times10^{20}$ cm$^{-2}$ (K km s$^{-1}$)$^{-1}$. The beam size and linear scale are shown at the bottom left and right respectively. Contours of $N$(H$_2$) are shown in red at levels of [1, 2, 5, 8, 10] $\times $ 10$^{21}$ cm$^{-2}$. The bottom panel shows the PDFs of $X$($^{12}$CO) and $X$($^{13}$CO).}  \label{fig:herschel_xfactors}
\end{figure}

\begin{figure}
    \centering
    \includegraphics[width=\linewidth]{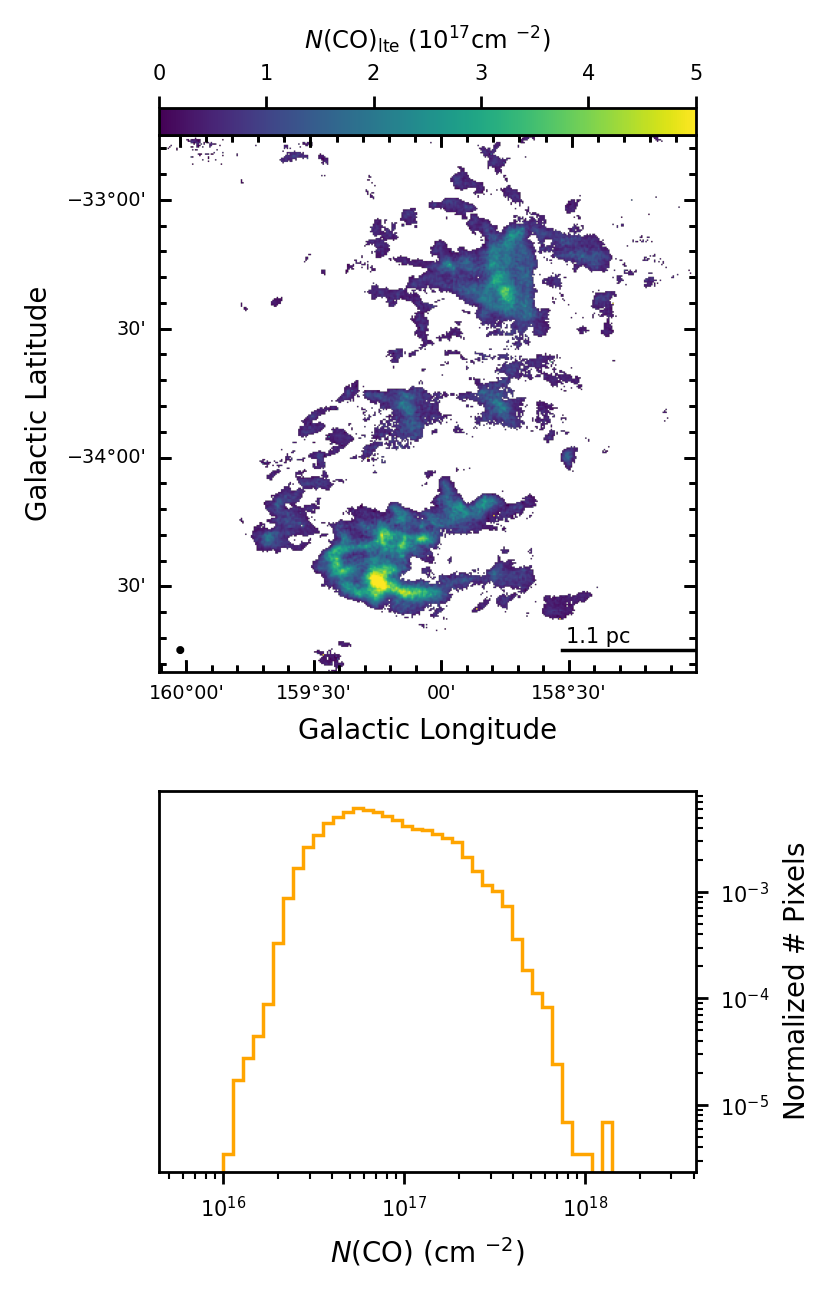}
    \caption{Carbon monoxide column density $N$(CO) for MBM12, derived from TRAO $^{12}$CO and $^{13}$CO $(J = 1-0)$ observations. The map shown on the top was estimated with standard LTE assumptions to derive $N$($^{13}$CO) and converted to total $N$(CO) with isotopologue ratio function from \citet{2014MNRAS.445.4055S}. The bottom panel shows the PDF for the $N$(CO) map.}
    \label{fig:co_column_density}
\end{figure}

\begin{figure}
    \centering
    \includegraphics[width=\linewidth]{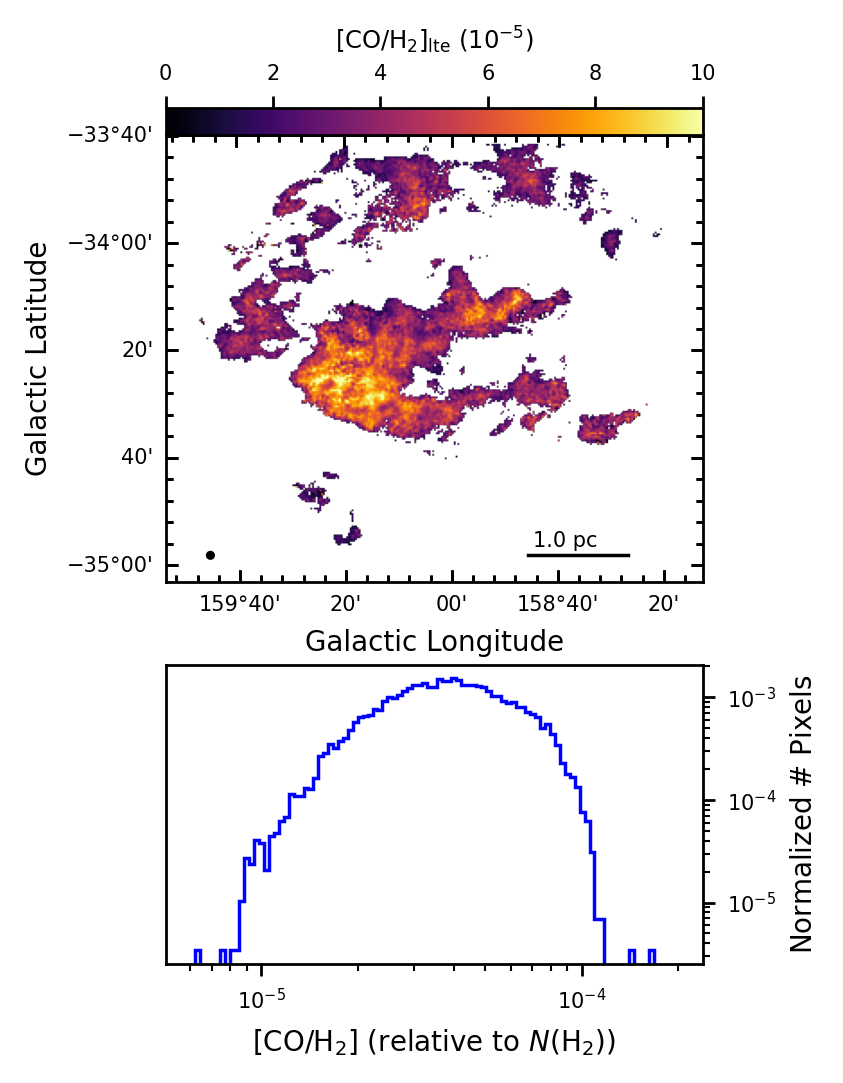}
    \caption{Map of CO abundance relative to H$_2$ column densities in the southern part of MBM12. The top panel shows $N$(CO) estimated from LTE  assumptions. The bottom panel shows the abundance PDF.}
    \label{fig:abundances}
\end{figure}

For MBM12, we show the relations between $N$(H$_2$), $N$(CO), $X$(CO) and [CO/H$_2$], and their spatial distributions and PDFs. Figure \ref{fig:environment_summary} shows a comparison of most of these quantities to each other. Figs. \ref{fig:nh2pdf} $-$ \ref{fig:abundances} show their spatial distributions and PDFs.

\noindent
The dust-opacity-calibrated $N$(H$_2$) maps of MBM12 range from $N$(H$_2$) $= 2\times10^{20}$ cm$^{-2}$ to $1.3\times10^{22}$ cm$^{-2}$. We generated a $N$(H$_2$) probability density function (PDF) (Fig. \ref{fig:nh2pdf}). The PDF shows multiple bumps, which may be the result of the superposition of several independent column density distributions in the field.

\noindent
We calculate $X$($^{12}$CO) and $X$($^{13}$CO) for the southern half of MBM12, by reprojecting the $40''$ {\it Herschel} $N$(H$_2$) map to the TRAO CO integrated intensity maps at $44''$ resolution (Fig. \ref{fig:herschel_xfactors}).
The morphology of $X$(CO) differs between the two isotopologues. In $^{12}$CO, the distribution is continuous, while in $^{13}$CO the Horseshoe is spatially disconnected from the Bow. The PDFs of $X$($^{12}$CO) and $X$($^{13}$CO) in Fig. \ref{fig:herschel_xfactors}\,c show double-peaked distributions connected by a powerlaw. We decompose the $X$(CO) PDFs into a lognormal at the peak with a truncated power law. The fitting equations and results of the fit are given in App. \ref{app:XCO-segmentation}. The combination of these distributions well described the $X$(CO) PDFs, except an excess at $X$($^{12}$CO) $= 10$ $X_{\rm Gal}$ and $X$($^{13}$CO) $= 40$ $X_{\rm Gal}$. Lognormal PDFs with deviations for $X$(CO) have been seen in $8.6'$ resolution maps by \citep[][App. B]{2022ApJ...931....9L}. We consider physical explanations for the $X$(CO) PDF in Sect. \ref{subsec:disc_xco}.

\noindent
The $N$(CO) varies in MBM12 from $1\times10^{16}$ cm$^{-2}$ up to $1\times10^{18}$ cm$^{-2}$ (Fig. \ref{fig:co_column_density}). The distribution of $N$(CO) shows filamentary structure in the Horseshoe. In North Compact, there are some peaks, but $N$(CO) is relatively constant throughout the structure.

\noindent
The abundances [CO/H$_2$] in MBM12 have an average of 5.9$\times 10^{-5}$. The spatial distributions of the [CO/H$_2$] have some interesting variation. At the centre of the Horseshoe, the abundances are between $4\times 10^{-5} - 1\times10^{-4}$, with a clumpy spatial distribution. The northern islands in the Bow show centrally peaked [CO/H$_2$] (Fig. \ref{fig:abundances}). The abundances are lognormal, centred at the average abundance. The abundance shows a power-law tail at the lower end.

\noindent
We compare our multi-tracer observables in Fig. \ref{fig:environment_summary}. Three main features are: (i) the quasi-linear dependence of $N$(CO) with $N$(H$_2$) (panel d) leading to an average uniform CO abundance [CO/H$_2$]=5.9$\times 10^{-5}$, (ii) at low $N$(H$_2$) (panel e), the two large excursions of $W$($^{12}$CO) below and above the average quasi-linear increase of $W$($^{12}$CO) with $N$(H$_2$), corresponding to $\langle X$($^{12}$CO)$\rangle=X_{\rm Gal}$. These one-dex excursions are responsible for the shape of the $X$($^{12}$CO) distribution with $N$(H$_2$) (panel j): the former, at the origin of the rise of $X$($^{12}$CO) above $X_{\rm Gal}$ at $N$(H$_2$) below 1.5 $\times 10^{21}$ cm$^{-2}$, is due to the drop of the CO abundance in the least UV-shielded layers while the latter, at the origin of all the $X$($^{12}$CO) values below $X_{\rm Gal}$, traces the fall-off of collisional de-excitation in low density gas, and (iii) the quasi-linear increase of $W$($^{13}$CO) with N(H$_2$) (panel f) leading to a $^{13}$CO integrated intensity close to 4 times weaker than that of $^{12}$CO, a result in agreement with early galactic CO and
$^{13}$CO line surveys that found that $^{12}$CO lines are $\sim$5 times brighter than $^{13}$CO lines \citep{stark1983comparison}.

\subsection{Multi-scale dynamics}
\label{subsec:multiscale_dynamics}
\begin{figure*}
    \centering
    \includegraphics[width=\linewidth]{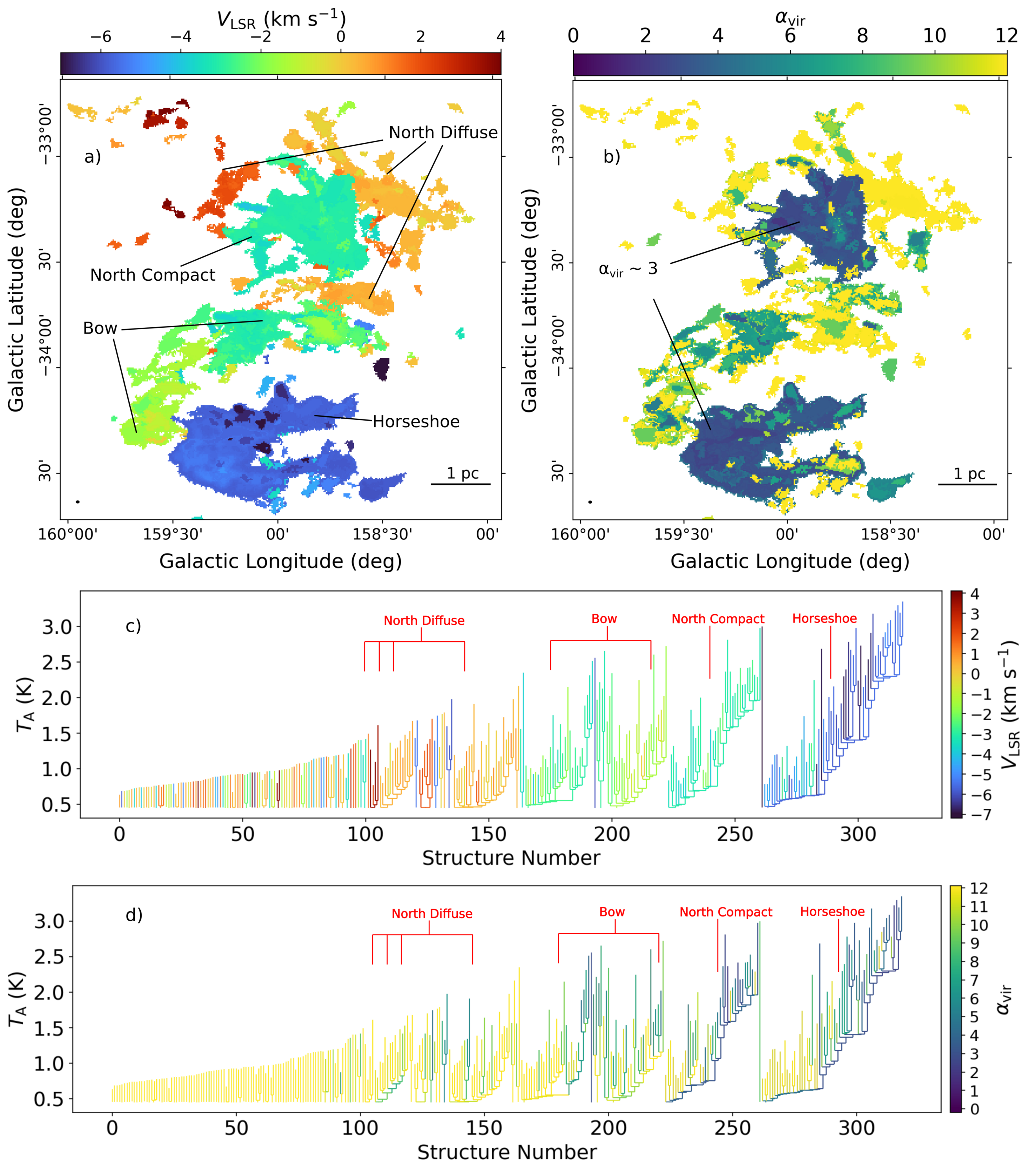}
    \caption{Hierarchical structure and virial estimates for MBM12 with $^{13}$CO ($J=1-0$) observations with the TRAO. Panel a: Centre velocities for dendrogram structures. The plot is made by plotting the largest scale structures first with a single colour corresponding to their $V_{\rm cen}$, and then the smaller scale structures are plotted on top. Panel b: Virial parameter estimates for dendrogram structures. The plot is made in the same way as panel a, but with the color defined by the value of $\alpha_{\rm vir}$. Panel c: Tree diagram for $^{13}$CO observations of MBM12. The tree diagram is coloured according to the value of $V_{\rm cen}$ for the structure with the same colour scale as panel a. The four largest-scale structures are shown in red. Panel d: Tree diagram for $\alpha_{\rm vir}$ estimated with the $^{13}$CO observations of MBM12. The tree diagram is coloured according to $\alpha_{\rm vir}$ with the same colour scale as panel b.}
    \label{fig:virial_dendro}
\end{figure*}
\subsubsection{How to interpret dendrograms}
The dendrogram algorithm is a useful way of segmenting PPV cubes into hierarchical structures. Our use of dendrograms assumes each voxel in PPV corresponds to a mass proportional to the voxel's $T_{\rm mb}$\citep{2008ApJ...679.1338R}. From Eq. $\ref{eq:virial_par}$ and the dendrogram algorithm, some considerations can be drawn. Higher in the hierarchy, the structures become smaller and have a higher average $T_{\rm mb}$. Therefore, one would generally expect $\alpha_{\rm vir}$ to decrease as one moves up the dendrogram hierarchy. A similar pattern should be expected for velocity dispersion. Therefore, looking only at the leaves of a dendrogram will systematically underestimate $\alpha_{\rm vir}$, as one ignores the pressure, shear and gravitational effects of the larger parent structure on $\alpha_{\rm vir}$ \citep{2006MNRAS.372..443B}. Therefore, it is helpful to look at the full hierarchy of $\alpha_{\rm vir}$ estimates. Virial parameters are calculated per structure, and are therefore difficult to visualize spatially. We produced a plot where we layer the contours of the structures, starting from the lowest hierarchical structure (i.e. the largest structure spatially). The contour is coloured by the value of the respective variable, either $V_{\rm LSR}$ or $\alpha_{\rm vir}$. We then successively layer the sub-structures on top of the largest structures to produce a two-dimensional visualization. This works well for simpler dendrograms, such as $^{13}$CO in MBM12 (Fig. \ref{fig:virial_dendro}). 
We colour the dendrogram tree diagrams with $V_{\rm LSR}$ and $\alpha_{\rm vir}$, and annotated the respective sub-structures (panels c and d of Figs \ref{fig:virial_dendro}). Virial parameters alone are not conclusive indicators of gravitational collapse, as they measure energy balance rather than stability. \citep{2006MNRAS.372..443B,2022MNRAS.517..885O,2023ASPC..534....1C}.
\subsubsection{Hierarchical structure in MBM12}
The $^{13}$CO emission is separated into the North-Diffuse, Bow, North-Compact and Horseshoe subregions (Fig. \ref{fig:virial_dendro}).

\noindent
The North Diffuse subregion has a ring shape in the north of MBM12, with most structures having $\alpha_{\rm vir} > 4$ \citep[also seen in][]{1990ApJ...351..165P}. The North Diffuse region has the same velocity as the small northern clumps at $V_{\rm LSR} \sim 4$ km s$^{-1}$.

\noindent 
The Bow subregion has two components in $^{13}$CO, east at $V_{\rm LSR} \sim -1$ km s$^{-1}$ and west at $-2.5$ km s$^{-1}$. This subregion is also detected with {\it Herschel}. The eastern Bow at $-1$ km s$^{-1}$ is likely denser than the western Bow, with $\alpha_{\rm vir} \lesssim 2.5$, and thin layers of large $X$($^{13}$CO) (middle panel of Fig. \ref{fig:herschel_xfactors}). The western Bow has two islands, the most western island with $\alpha_{\rm vir} > 5$, and an eastern island with $\alpha_{\rm vir} \sim 2$. The western Bow has thicker layers of large $X$($^{13}$CO) (middle panel of Fig. \ref{fig:herschel_xfactors}) than the eastern Bow. 

\noindent The North Compact subregion has complex hierarchical structure. It was not covered with {\it Herschel}. It has some sub-structures with $\alpha_{\rm vir} > 5$ (panel d of Fig. \ref{fig:virial_dendro}), but the whole structure has $\alpha_{\rm vir} \sim 3$. The region has the largest $N$(CO) in the north of MBM12 (Fig. \ref{fig:co_column_density}) but is not strongly centrally peaked. There are single structures in this subregion (Fig. \ref{fig:dendro_scale_dependence}) with $\alpha_{\rm vir} < 2$, but most of this structure has $\alpha_{\rm vir} > 2$.

\noindent The Horseshoe, at $V_{\rm LSR} = -5.5$ km s$^{-1}$ has the most complex hierarchical structure in MBM12. Some structures at the southern part of the Horseshoe have $\alpha_{\rm vir} > 5$ (Fig. \ref{fig:virial_dendro} b). Some of these may be fore- or background clumps, as they are also at different velocities. The Horseshoe has the largest average $N$(H$_2$) in MBM12 and has $\alpha_{\rm vir} \sim 3$.
\subsubsection{Scale dependence of $\alpha_{\rm vir}$}
\label{subsubsec:virial_scale_dependance}
\label{subsubsec:alpha_scale_dependence}
\begin{figure*}
    \centering
    \includegraphics[width=\linewidth]{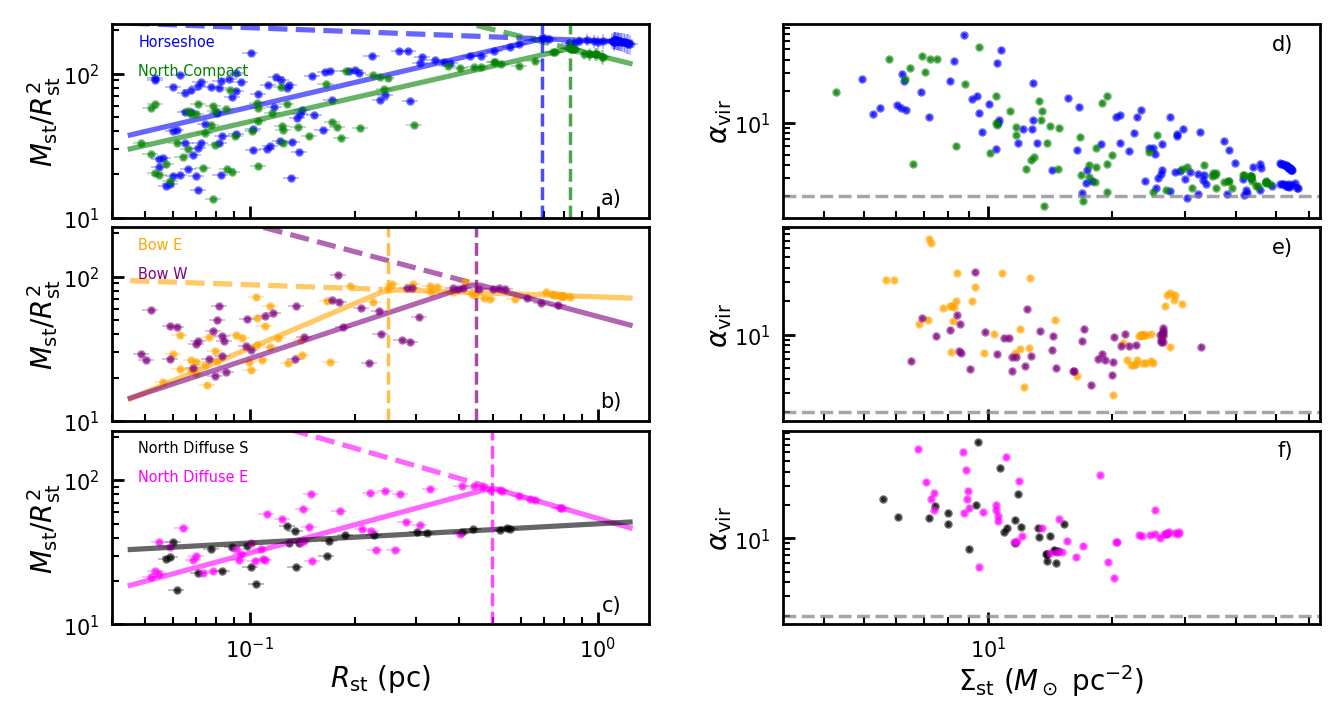}
    \caption{Scale dependence of $^{13}$CO dendrogram structures in MBM12. The colors are for the Horseshoe (blue), North Compact (green), eastern Bow (yellow), western Bow (purple), southern North Diffuse (black) and eastern North Diffuse (grey). Broken power law fits are also shown, with the respective fit values in Tab. \ref{tab:mass-radius-fits}. Vertical lines indicate the power law breaking point. The vertical axis of panels a) - c) ($M_{\rm st}/R_{\rm st}^2$) is in units of $M_\odot$ pc$^{-2}$.}
    \label{fig:dendro_scale_dependence}
\end{figure*}
We examine the interdependence of $M_{\rm st}$ and $\alpha_{\rm vir}$ on $R_{\rm st}$ for the largest structures identified in the $^{13}$CO dendrogram (Fig. \ref{fig:virial_dendro}c). We fitted a broken power law to the masses $M_{\rm st}$ and radii $R_{\rm st}$ of $^{13}$CO dendrogram structures in MBM12. The functional form is
\begin{equation}
    \Bigg(\frac{M_{\rm st}}{1 M_\odot}\Bigg) = A_i \Bigg(\frac{R_{\rm st}}{1 {\rm pc}}\Bigg)^{\alpha_i}
    \label{eq:brokenplaw}
\end{equation}
where $i = 1$ if $R_{\rm st} < R_{\rm break}$ and $i = 2$ if $R_{\rm st} > R_{\rm break}$. The best fit parameters with uncertainties are shown in Tab. \ref{tab:mass-radius-fits}. To ensure continuity between the two power laws, from Eq. \ref{eq:brokenplaw} the break-point $R_{\rm break}$ can be calculated directly:
\begin{equation}
    R_{\rm break} = \Bigg(\frac{A_1}{A_2}\Bigg)^{\frac{1}{\alpha_2 - \alpha_1}}
\end{equation}
\begin{table*}
    \centering
    \caption{Fitted parameters for the broken power law fits to the mass-size relations (Fig. \ref{fig:dendro_scale_dependence}).}
    \begin{tabular}{llllllll}
    \hline
     ID & Subregion & $A_1$ & $\alpha_1$ & $A_2$ & $\alpha_2$ & $R_{\rm break}$ (pc) \\
    \hline
    1 & Horseshoe & $215\pm8$ & $2.56\pm0.06$ & $169\pm1$ & $1.91\pm0.02$ & $0.69$ \\
    2 & North Compact & $165\pm2$ & $2.55\pm0.03$ & $134\pm1$ & $1.40\pm0.04$ & $0.83$ \\
    3 & Bow E & $339\pm312$ & $3.02\pm0.55$ & $73\pm1$ & $1.92\pm0.03$ & $0.25$ \\
    4 & Bow W & $170\pm28$ & $2.80\pm0.16$ & $53\pm2$ & $1.36\pm0.07$ & $0.45$ \\
    5 & North Diffuse S & $50\pm1$ & $2.13\pm0.02$ & -- & -- & -- \\
    6 & North Diffuse E & $139\pm19$ & $2.65\pm0.16$ & $54\pm2$ & $1.30\pm0.09$ & $0.50$ \\
    \hline
    \end{tabular}
    \tablefoot{The parameters ($A_1$,$\alpha_1$) were for $R_{\rm st} < R_{\rm break}$, while ($A_2$,$\alpha_2$) was for $R_{\rm st} > R_{\rm break}$ North Diffuse S is fit with a single power law.}
    \label{tab:mass-radius-fits}
\end{table*} From the dendrogram we take trunk-level structures that have more than eleven sub-structures. Spatial investigation reveals that these structures are the horseshoe, North Compact, eastern Bow, western Bow, southern North Diffuse, and eastern North Diffuse. We then plot the mass-size relation normalized by $R_{\rm st}^2$, and the surface density dependence of $\alpha_{\rm vir}$ for all hierarchical structures (Fig. \ref{fig:dendro_scale_dependence}). Note that these data points are not independent, as they share voxels in PPV. Therefore we are rather estimating the slope of the change in mass with radius, comparable to the differential virial analysis of \citep{2025OJAp....8E..91K}. We found correlations between mass and radius, that are best described by a broken power laws (Fig. \ref{fig:dendro_scale_dependence} a). Tab. \ref{tab:mass-radius-fits} show the scaling parameters $A_i$, power law slopes $\alpha_i$ and break points $R_{\rm break}$ for all the substructures. The large scale ($R_{\rm st} > R_{\rm break}$) indices were close to $M \propto R^2$ for the Horseshoe and eastern Bow with $\alpha_2 = 1.91\pm0.02$ and $\alpha=1.92 \pm 0.03$ respectively. For the other subregions, the outer power law index is sub-Larson, with $\alpha_2 < 1.5$. The outer scaling factor $A_2$ varies between $54-169$, with the largest value at the Horseshoe, and the smallest at North Diffuse subregions. However, the power law index is super-Larson for all low-mass subregions in the inner power law, with $\alpha_1 > 2$, and the statistically significant scaling factors $A_1$ varying between 50$-215$. The virial parameter is increasing with decreasing $\Sigma_{\rm st}$. The North Compact is the only region which contained $\alpha_{\rm vir} < 2$ structures. However, $\alpha_{\rm vir}$ has some variation depending on the scale at which the structure is analysed. Many of the smallest scale structures in the Horseshoe and North Compact have large $\alpha_{\rm vir} > 2$. The other structures have $\alpha_{\rm vir} \sim8$ at parsec scales (Fig. \ref{fig:dendro_scale_dependence} b), with variability in $\alpha_{\rm vir}$ towards smaller scales. No structures in eastern Bow, western Bow, southern North Diffuse, eastern North Diffuse have $\alpha_{\rm vir} \sim 2$ at any scale. 
\subsection{Histograms of relative orientations}
\label{subsec:results_magnetic_fieldsMBM12}
\begin{figure*}
    \centering
    \includegraphics[width=\linewidth]{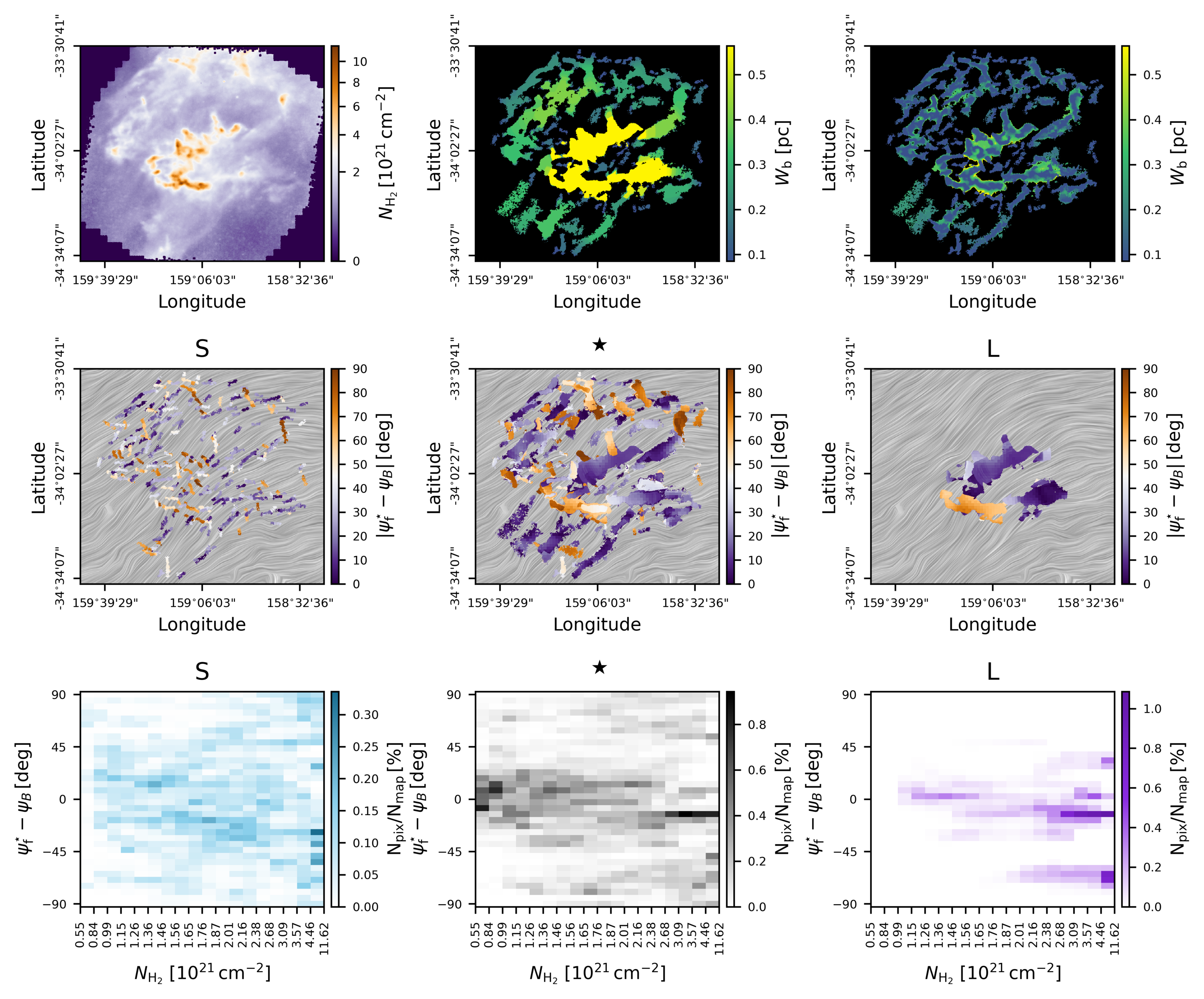}
    \caption{Top left: MBM12 H$_2$ column density map. Top middle: Network of filaments reconstructed with {\tt FilDReaMS}, with the largest bar width, $W_{\rm b}$, per pixel shown. Top right: Same network of filaments with the smallest $W_{\rm b}$ per pixel shown. Middle left, middle and right: Map of the relative orientation for the smallest $W_{\rm b}$ filaments, most significant filaments across $W_{\rm b}$ and largest $W_{\rm b}$ filaments respectively. \textbf{\textit{B}$_{\rm{PoS}}$} orientation is visualised using line integral convolution (LIC) in grayscale in the background. Bottom left, medium and right: Histograms of relative orientations as a function of $N$(H$_2$) for the smallest $W_{\rm b}$ filaments, most significant filaments across $W_{\rm b}$ and largest $W_{\rm b}$ filaments respectively.}
    \label{fig:fildreams_hro}
\end{figure*}
We examine the relative orientation between $N$(H$_2$) structures and the plane-of-sky magnetic field \textbf{\textit{B}$_{\rm{PoS}}$}. Figure~\ref{fig:fildreams_hro} summarizes our histogram of relative orientation analysis in MBM12. The top left panel shows the {\it Herschel} $N$(H$_2$) map, from which networks of filaments were extracted. These networks are shown in the top middle and top right panels, with the largest-$W_{\rm b}$ and smallest-$W_{\rm b}$ filaments per pixel shown. These figures reveal very thin and low-$N$(H$_2$) striations around the Horseshoe and Bow subregions. The Horseshoe has significantly thicker filaments than the Bow, with bar widths $\sim 0.6$ pc, with a finer, continuous structure, while the Bow appears more disconnected, with bar widths of $\sim 0.4$ pc. The middle row of Fig.~\ref{fig:fildreams_hro} shows maps of the relative orientation between filaments and \textbf{\textit{B}$_{\rm{PoS}}$} for the smallest-$W_{\rm b}$ filaments (left panel), largest-$W_{\rm b}$ filaments (right panel), and most significant filaments across all $W_{\rm b}$ (center panel). The bottom row of Fig.~\ref{fig:fildreams_hro} shows the histograms of relative orientations as a function of $N$(H$_2$) for the same sets of filaments. At small scales, most filaments are roughly parallel to \textbf{\textit{B}$_{\rm{PoS}}$} at low $N$(H$_2$), with $\lvert \psi_{\rm f} - \psi_B \rvert \lesssim 35^{\circ}$ for $N_{\rm H_2} \leq 4.5 \times 10^{21}\,\rm{cm^{-2}}$, while high-$N$(H$_2$) filaments have more random orientations, with a small preference towards $-45^{\circ}$. The situation is clearer for the most significant and largest-$W_{\rm b}$ filaments. Similarly, most filaments at low-$N$(H$_2$) are roughly parallel to \textbf{\textit{B}$_{\rm{PoS}}$} while the high-$N$(H$_2$) filaments, which correspond to the Horseshoe, shows that the northern part is mostly parallel to \textbf{\textit{B}$_{\rm{PoS}}$} and the southern part is roughly perpendicular to \textbf{\textit{B}$_{\rm{PoS}}$}. The HROs in MBM12 seem to indicate that at low-$N$(H$_2$) filaments tend to be smaller and oriented parallel to the magnetic field. As the $N$(H$_2$) increases, the filament size also increases, and these large scale filaments are either parallel or perpendicular to the magnetic field. At high $N$(H$_2$), structures are located in the southern part of the Horseshoe, perpendicular to the magnetic field, where both the large scale filaments are co-spatial in the plane of the sky.
\section{Discussion}
\label{sec:discussion}
\subsection{Dust opacity variation and $N$(H$_2$) estimation}
\begin{table*}
    \centering
    \caption{Summary of $\kappa_{\rm 1200}$ estimating strategies and the consequent values.}
    \begin{tabular}{lllll}
    \hline
        No. & Method & $\kappa_{\rm 1200}$ ($\beta = 2$) & Range in $\kappa_{\rm 1200}$ & References \\
        & & (g$^{-1}$ cm$^{2}$) & (g$^{-1}$ cm$^{2}$) & \\
    \hline
   1 & Common Literature & 0.1 & N$/$A & \citep{1983QJRAS..24..267H}\\
   2 & Common Literature & 0.144 & 0.136 $-$ 0.149\tablefootmark{a} & \citep{1990AJ.....99..924B}\\
   3 & Survey &  0.093 $\pm$ 0.016 & 0.065  $-$ 0.12\tablefootmark{b} & \citep{2022ApJ...931....9L} \\
   4 &  Survey & 0.112$\pm 0.014$ & 0.084 $-$ 0.280 & {\citep{2015A&A...584A..93J}} \\
   5 & Calibrator $N$(H) & 0.124$\pm 0.006$ & 0.124 $-$ 0.176\tablefootmark{c} & {\citep{2017A&A...601A..78R}} \\
   6 & Calibrator $N$(H) & 0.192$\pm$0.007 & $0.16 - 0.24$\tablefootmark{d} & Extinction with NICEST (Sect. \ref{subsec:nh2})\\
    \hline
    \end{tabular}
    \tablefoot{For the determinations of the values and ranges, see App. \ref{app:dust_opacity_determination}. \tablefoottext{a}{For $\beta = 1.7 - 2.2$.} \tablefoottext{b}{For $\beta = 1.7 - 2.2$ between 353 GHz and 1200 GHz.} \tablefoottext{c}{For $f_{\rm mol} = 0.7 - 1$.} \tablefoottext{d}{Spatial variation.}}
    \label{tab:dust_opacity_variation}
\end{table*}
The dust opacity $\kappa_\nu$ is a function of various dust properties (size distribution, chemical composition and structure). Dust models can predict $\kappa_\nu$ among other things \citep[e.g.][]{2023ApJ...948...55H,2024A&A...684A..34Y}. It is known that the far-infrared dust opacity varies in the diffuse ISM, with less pronounced variations in NIR extinction \citep{2014A&A...566A..55P,2015ApJ...811..118R,2018ApJ...862...49N}. Between the diffuse ISM and molecular clouds, there is a factor of three change in the ratio of  the 250 $\mu$m and NIR opacities \citep{2015A&A...584A..93J}. Empirical $\kappa_\nu$ calibration is important as column density estimates propagate into mass, density and abundance estimates. We summarise some methods for estimating $\kappa_{\rm 1200}$ in Table \ref{tab:dust_opacity_variation} (see App. \ref{app:dust_opacity_determination}).
We recommend the $\kappa_{\nu}$ calibration method of \citet{2022ApJ...931....9L} with NIR extinction. Even if NIR extinction maps do not have high resolution, they can give a cloud averaged $\kappa_\nu$, or spatial variation in some cases that is better than assuming a constant from literature. The NIR extinction to column density ratio is not necessarily constant if $R_{\rm V}$ varies, but the empirical models of \citet{2023ApJ...948...55H} can be used to quantify the uncertainty of $R_{\rm V}$. Informed use of $\kappa_\nu$ would remove bias in masses and chemical abundances in diverse star forming environments.
\subsection{Intra-cloud $X$(CO) variations}
\label{subsec:disc_xco}
The $X$(CO) factor is a measure of H$_2$ molecules per CO emission. On theoretical grounds we expect it to be affected by CO abundances, H$_2$ kinetic temperature, H$_2$ density, dust temperature and the surrounding radiation field \citep{2011MNRAS.415.3253S,2021ApJ...908...76L}. Molecular H$_2$ and CO are photodissociated in the interstellar radiation field if shielding is too weak \citep{1988ApJ...334..771V,2018ApJ...858...16G}. Shielding can occur due to self-shielding, shielding by other molecules or attenuation by dust \citep{1988ApJ...334..771V}. Molecular hydrogen forms at low densities ($> 30$ cm$^{-3}$) with efficient self shielding, while CO requires dust to attenuate the radiation and therefore forms at hydrogen densities $> 100$ cm$^{-3}$ \citep{1988ApJ...334..771V,2010MNRAS.404....2G,2018ApJ...858...16G}. In these intermediate densities $30 - 100$ cm$^{-3}$ is CO dark gas, which would lead one to expect $X$(CO) $\rightarrow \infty$ as CO starts to form at the outer edges of molecular clouds. However observations of diffuse clouds have rather found $X$(CO) $\leq X_{\rm Gal}$ \citep{2010A&A...518A..45L,2012A&A...541A..58L}. This can be explained by low collisional de-excitation for CO at low densities, where CO is more effective per H$_2$ molecule at emission \citep{2012A&ARv..20...55H}. The variance of the $X$(CO) factor is also very high in diffuse regions, where $n_{\rm H_2} < 400$ cm$^{-3}$ \citep{2011MNRAS.415.3253S}. Deeper into the cloud, where $A_{\rm V}$ is between 1 and 2 mag, the dust shields the interstellar radiation field (ISRF) for the formation of CO, $X$(CO) exponentially decreases \citep{2011MNRAS.415.3253S,2016MNRAS.460...82S}. However, at the largest densities, the lines may self-absorb, or CO is frozen onto dust grains, which then increases $X$(CO) again \citep{2010ApJ...721..686P,2016MNRAS.460...82S}. We see these regimes in Fig. \ref{fig:environment_summary} j and k. For $N$(H$_2$) below $10^{21}$ cm$^{-2}$, we have high $X$($^{12}$CO) values, but also many where $X$($^{12}$CO) $< X_{\rm Gal}$. The mean value of $X$($^{12}$CO) was very close to the galactic average recommended by \citet{2013ARA&A..51..207B}. Our estimate is twice the value of $\approx 0.5 X_{\rm Gal}$ surrounding MBM12 from \citet{2017A&A...601A..78R}. In Perseus $X$($^{12}$CO) was measured as $0.5 - 1.5 X_{\rm Gal}$ \citep{2008ApJ...679..481P}. The relationship between $X$(CO) and $N$(H$_2$) has been investigated theoretically and through observations of the California MC as well \citep{2015ApJ...805...58K,2016MNRAS.460...82S,2021ApJ...908...76L}. 
\citet{1988ApJ...326..909M} calculated $X$(CO) for various high-latitude clouds (excluding MBM12) and found an average value of $X$(CO) $= 3.2\pm 0.6$ $X_{\rm Gal}$ for their full sample, or 1.6 $X_{\rm Gal}$ if they excluded some dark clouds. They also estimated abundances of $4-12 \times10^{-5}$ which is closer to our estimate of MBM12 than to the average Milky Way value.  \citet{2013MNRAS.436.1152C} estimated $X$(CO) $= 0.65$ $X_{\rm Gal}$ for MBM 40. In MBM16, $X$(CO) $= 3.6$ $X_{\rm Gal}$ and in MBM40 $X$(CO) $=$ 1.3 $X_{\rm Gal}$ \citep{1998ApJ...504..290M}. The $X$(CO) PDFs possibly encode this chemical behaviour (Fig. \ref{fig:herschel_xfactors}). Close to the average value of $X$(CO) the PDF was lognormal, with a similar standard deviation of around 1.4 $X_{\rm Gal}$. In App \ref{app:XCO-segmentation}, we use the statistical behaviour of the $X$(CO) PDF to visualize the physical transitions in MBM12 Fig. \ref{fig:xco_segmentation}. If other molecular clouds show similar lognormal and powerlaw behaviour in their $X$(CO) PDFs, the PDF properties can be used as a measure of chemistry and excitation. Most works only focus on $X$(CO) as a conversion factor for mass. However, our $44''$ investigation of $X$(CO) implies that it may be a useful probe of molecular cloud substructure in its own right. Radiative transfer modelling of the $X$(CO) PDFs physical dependencies combined with estimates from more sources may reveal if $X$(CO) PDFs have any diagnostic potential.
\subsection{CO abundances}
The highest CO column densities in MBM12 was $1 \times 10^{18}$ cm$^{-2}$. The high column densities are associated with the Horseshoe and North Compact (Fig. \ref{fig:co_column_density}). The measured $N$(CO) for MBM12 of a few 10$^{17}$ is higher than any MBM clouds in the sample of \citet{1991ApJ...366..141V}. The mean [CO/H$_2$] of $5.91\times 10^{-5}$ of MBM12 is lower than the canonical $10^{-4}$ in Milky Way clouds such as Taurus and Orion B \citep{2010ApJ...721..686P,2013ARA&A..51..207B,2021A&A...645A..26R}. Low abundances like this have been observed before \citep{2007ApJ...658..446B,2013ApJ...775L...2L,2023ApJ...942..101L}. The methodological assumptions that go into abundance calculations ($\kappa_\nu$ for $N$(H$_2$), constant $T_{\rm ex}$, isotope ratio, LTE only vs non-LTE) make precise comparison with other works difficult. In the case of Taurus \citet{2010ApJ...721..686P}, assumed a constant isotope ratio of 69, which increases their $N$(CO) estimate by up to 70$\%$, compared to our use of a variable isotope ratio \citep{2014MNRAS.445.4055S}. Variable isotope ratio would also change the shape of the $N$(CO) and [CO/H$_2$] PDFs. The environmental differences between Taurus and MBM12 play a role in the abundance differences. 
\subsection{Virial analysis in MBM12}
\label{subsubsec:disc_virial_analysis_mbm12}
We used virial analysis as one tool among many for assessing virial equilibrium. In App. \ref{app:virial_analysis_interpretation} we discuss some of the historical ambiguities in the interpretation of $\alpha_{\rm vir}$. We find $\alpha_{\rm vir} \gtrsim 3$ for all regions in MBM12. In all cases, $\alpha_{\rm vir}$ did not vary significantly with scale. The scale independence of $\alpha_{\rm vir}$ has also been seen in other regions, including Serpens South and NGC 253 \citep{2024RAA....24f5003L,2024ApJ...969...70F,2025ApJ...993..193O}. Scale independent $\alpha_{\rm vir}$ is expected for structures. If the structures follow Larson's laws, $M_{\rm st} \propto R^2_{\rm st}$ and $\sigma_{\rm v, st} \propto R_{\rm st}^{0.5}$, then from Eq. \ref{eq:virial_par} $\alpha_{\rm vir} \propto \sigma_{\rm st}^{2}R_{\rm st}M_{\rm st}^{-1} \propto 1$. The scaling factors in front of the mass-size relations (Fig. \ref{fig:dendro_scale_dependence}) in the Horseshoe are up to three times higher than that of the northern regions. 
The $M_{\rm st} \propto R_{\rm st}^2$ and $\sigma_{\rm v, st} \propto R_{\rm st}^{0.5}$ hierarchy is the locus of the gravitational instability threshold of self-gravitating polytropes of increasing temperature immersed in a common pressure reservoir $P_{\rm ext}$. The structures of this hierarchy are the result of fragmentation of the larger scales, which occurs prior to collapse, and behave as a N-body system in virial equilibrium \citep{1987A&A...171..225C}. The interesting point here is that the pre-factor of the mass-size relation is proportional to $P_{\rm ext} ^{0.5}$ which shows that the external pressure of the structures of lowest $\alpha_{\rm vir}$ is about 10 times larger than that of the structures of largest $\alpha_{\rm vir}$. For a turbulent core with a constant surface density, an increased internal pressure (which is related to the external pressure) should increase $\alpha_{\rm vir}$ \citep{1999ApJ...522..313M,2003ApJ...585..850M}. We do observe that the smaller scale structures $R_{\rm st} < R_{\rm break}$, with larger mass-size coefficients have larger virial parameters (Fig. \ref{fig:dendro_scale_dependence}\,d). Additional investigation will be required into the role of pressure in the scale dependent behaviour of $\alpha_{\rm vir}$, such as the analysis proposed by \citet{2025OJAp....8E..91K}. 
\subsection{Magnetic fields and gravitational stability in MBM12}
\label{subsec:bfields_in_mbm12}
From the analysis of ideal MHD equations, \citet{2017A&A...607A...2S} showed that perpendicular and parallel alignments between magnetic field orientations and column density structures are attractors in the equations of motion. In strongly magnetized (plasma $\beta \sim 0.1$) gas the transition between these two attractor modes are induced through convergent flows where $\nabla\cdot \vec{\rm {v}} < 0$ \citep{2013ApJ...774..128S,2017A&A...607A...2S}. Convergent flows can either be large scale flows or self-gravity which dominates magnetic pressure gradients in molecular clouds \citep{2016ApJ...829...84C}. Observational studies of ten Gould Belt regions with {\it Planck} dust emission and polarisation measurements found a transition column density $N_{\parallel\rightarrow\perp}$(H) $= 5\times 10^{21}$ cm$^{-2}$ \citep{2016A&A...586A.138P}. This transition $N$(H) is close to the $N_{\rm c\rightarrow pl}$(H) $= 10^{21}$ cm$^{-2}$ where the upper limits of magnetic field strengths from Zeeman measurements transition from constant to power law \citep{2012ARA&A..50...29C,2017A&A...607A...2S}. The connection between the $N_{\parallel\rightarrow\perp}$ to $N_{\rm c\rightarrow pl}$(H) may indicate a shared mechanism of super-critical filaments between these two transitions. However, this interpretation is weakened by the large scatter in measured $N_{\parallel\rightarrow\perp}$(H), which ranges from $2\times 10^{21}$ cm$^{-2}$ to $23 \times 10^{21}$ cm$^{-2}$, or many clouds not displaying a transition at all \citep{2016A&A...586A.138P,2017A&A...603A..64S,2022A&A...668A..42C}. MBM12 displays a transition column density of $N_{\parallel\rightarrow\perp}({\rm H_2}) \sim 4.5 \times 10^{21}$ cm$^{-2}$ and $\alpha \simeq 3$ in the Horseshoe. Our virial parameter ignores contributions from internal magnetic pressure or external thermal pressure, which brings up the question of whether the magnetic field is supporting MBM12 from collapse. For a single source, a parallel to perpendicular transition is difficult to interpret \citep[Sect. 6.1.3 of][]{2023ASPC..534..193P}. The presence of the relative orientation transition indicates the presence of strong magnetic fields $\beta < 1$, convergent velocity flows and possibly super-Alfvénic gas motions. The south of MBM12 Horseshoe would therefore be a candidate for higher resolution kinematic follow-up, to assess whether there are signs of infall or gravitational collapse. 
\section{Conclusion}
We report large-scale $^{12}$CO and $^{13}$CO ($J=1-0$) line observations of MBM12 with the TRAO radio telescope combined with {\it Herschel} and {\it Planck} column density estimates. For the first time in MBM12, we combined an analysis of molecular gas properties ($N$(H$_2$), $X$(CO), $N$(CO), [CO/H$_2$]), dynamics (multi-scale virial parameters $\alpha_{\rm vir}$) and magnetic fields (with histogram of relative orientations). 
\begin{enumerate}
    \item We estimated the spatial variation in the molecular gas fraction and 1200 GHz dust opacity, with $\langle\kappa_{1200}\rangle = (192 \pm 7)\times10^{-3}$ cm$^2$ g$^{-1}$, resulting in column densities that range from $2\times10^{20}$ cm$^{-2}$ to $1.3\times 10^{22}$ cm$^{-2}$.
    \item The average CO-to-H$_2$ conversion factors were $X$($^{12}$CO) $= 2 \times 10^{20}$ cm$^{-2}$ (K km s$^{-1}$)$^{-1}$ and $X$($^{13}$CO) $= 7.8 \times 10^{20}$ cm$^{-2}$ (K km s$^{-1}$)$^{-1}$ with significant variation across the field. The average $X$($^{12}$CO) value in MBM12 is equal to the galactic average. We decomposed the $X$(CO) PDFs into lognormal and power law components, with the power law slopes of $X$($^{12}$CO) $=-1.21\pm0.03$ and $X$($^{13}$CO) $= -0.78\pm 0.04$. Future work will investigate the parameters which influence the statistical intracloud behaviour of $X$(CO).
    \item We estimated the CO column density $N$(CO), and  the CO abundances [CO/H$_2$]. We incorporated variable $^{12}$CO/$^{13}$CO isotope ratios calibrated by simulations. We measured $N$(CO) ranging from $1\times10^{16}$ cm$^{-2}$ to $1\times 10^{18}$ cm$^{-2}$. The abundances [CO/H$_2$] ranged from $0.9\times 10^{-5}$ to $1\times 10^{-4}$ with an average of 5.9$\times 10^{-5}$. 
    \item We used dendrograms to calculate multi-scale virial parameters $\alpha_{\rm vir}$ for MBM12. For the Horseshoe and North Compact subregions, for most structures $\alpha_{\rm vir} \simeq 3$ at scales of 0.05$-$ 1 pc with masses scaling $M\propto R^{2}$ and Larson scaling $\sigma_{\rm v} \propto R^{0.5}$. On the other hand for the North Diffuse and Bow regions $\alpha_{\rm vir} \simeq 8$ at scales of 0.05$-$ 1 pc. The Horseshoe has a mass-radius scaling coefficient three times larger than the other subregions, indicating an order of magnitude stronger external pressure on the Horseshoe, compared to the other subregions. Scale dependent $\alpha_{\rm vir}$ measurements can highlight the role of external pressure in regulating star formation molecular clouds.
    \item We estimated the histogram of relative orientations in MBM12, and found column density structures parallel to the magnetic field orientation for $N({\rm H_2}) < 4.5 \times 10^{21}$ cm$^{-2}$, and some perpendicular structures thereafter. Perpendicular structures were only detected in the south of the Horseshoe subregion, and may indicate the presence of strong magnetic fields and convergent velocity flows.
\end{enumerate}
Our first case study of MBM12 in the B-FROST survey highlights the complementary value of tracing environment, dynamics and magnetic fields. We also stress the value of empirical $\kappa_\nu$ calibration for H$_2$ column densities and assumptions about LTE and isotope ratios for CO column densities. The multi-tracer approach in MBM12 enables comparative studies of low- vs high-mass star formation under varying conditions of turbulence, feedback, and magnetization in large scale molecular cloud surveys.
\begin{acknowledgements}
We thank Quentin Remy for supplying the opacity maps of anti-centre clouds. JMV and MJ acknowledge support from the Academy of Finland grant No 348342. EM is funded by the University of Helsinki doctoral school in particle physics and universe sciences (PAPU). VMP acknowledges financial support by the grant PID2020-115892GB-I00, funded by MCIN/AEI/10.13039/501100011033 and by
the grant CEX2019-000918-M funded by MCIN/AEI/10.13039/501100011033. VMP gratefully acknowledges also financial support from the European Research Council via the ERC Synergy Grant "ECOGAL" (project ID 855130). DA acknowledges NU FDCRGP No201223FD8821. We acknowledge the use of data provided by the Centre d'Analyse de Données Etendues (CADE), a service of IRAP-UPS/CNRS  \citep[http://cade.irap.omp.eu,][]{2012A&A...543A.103P}. C.W.L is supported by the Basic Science Research Program through the NRF funded by the Ministry of Education, Science and Technology (grant No. NRF- 2019R1A2C1010851) and by the Korea Astronomy and Space Science Institute grant funded by the Korea government (MSIT; project No. 2025-1-841-02). This research made use of astrodendro, a Python package to compute dendrograms of Astronomical data (http://www.dendrograms.org/). This work made use of Astropy:\footnote{http://www.astropy.org} a community-developed core Python package and an ecosystem of tools and resources for astronomy \citep{astropy:2013, astropy:2018, astropy:2022}.

\end{acknowledgements}
\bibliographystyle{aa} 
\bibliography{BFROSTII} 
\begin{appendix}
\section{Polarisation angles}\label{app:polarization_angle}
We derived the plane-of-sky magnetic field (\textbf{\textit{B}$_{\rm{PoS}}$}) orientation angle, $\psi_{B}$, from linear polarisation components $Q$ and $U$ using
\begin{equation}\label{eq:Borientation}
    \psi_{B} = \frac{1}{2} \ \rm{arctan}\left(\frac{{\it U}}{{\it Q}}\right) \pm 90^{\circ} \ ,
\end{equation}
which corresponds to the linear polarisation position angle rotated by $90^{\circ}$, and where $\arctan$ is the two-argument arctangent function defined from $-180^\circ$ to $+180^\circ$. Following \cite{Montier2015}, we computed the uncertainty in the linear polarisation position angle, $\sigma_{\psi_{B}}$ using
\begin{equation}\label{eq:Borientationuncertainty}
    \sigma_{\psi_{B}} = \frac{1}{2} \sqrt{\frac{Q^2 \sigma_{\rm{U}}^2 + U^2 \sigma_{\rm{Q}}^2 - 2QU \sigma_{\rm{QU}}}{(Q^2 + U^2)^2}} \ ,
\end{equation}
with $\sigma_{\rm{Q}}$ and $\sigma_{\rm{U}}$ the respective uncertainties in $Q$ and $U$, and $\sigma_{\rm{QU}}$ the covariance of $Q$ and $U$. We adopt the IAU convention, where the linear polarisation position angle increases from Galactic north to the east (opposite to the {\tt Healpix} convention requiring the flipping of the sign of \textit{U} in data taken from the \textit{Planck} Legacy Archive). There is a 180$^{\circ}$ ambiguity when estimating $\psi_{B}$ from linear polarisation that only gives the magnetic field orientation. However, it is irrelevant in the study of relative orientations, thus we can define $\psi_{B}$ in the range [$-90^{\circ}$, $+90^{\circ}$] from Eq.~\eqref{eq:Borientation} by choosing the sign in the last term accordingly.
\section{Spatial variation in line ratios}
\begin{figure*}
    \centering
    \includegraphics[width=0.8\linewidth]{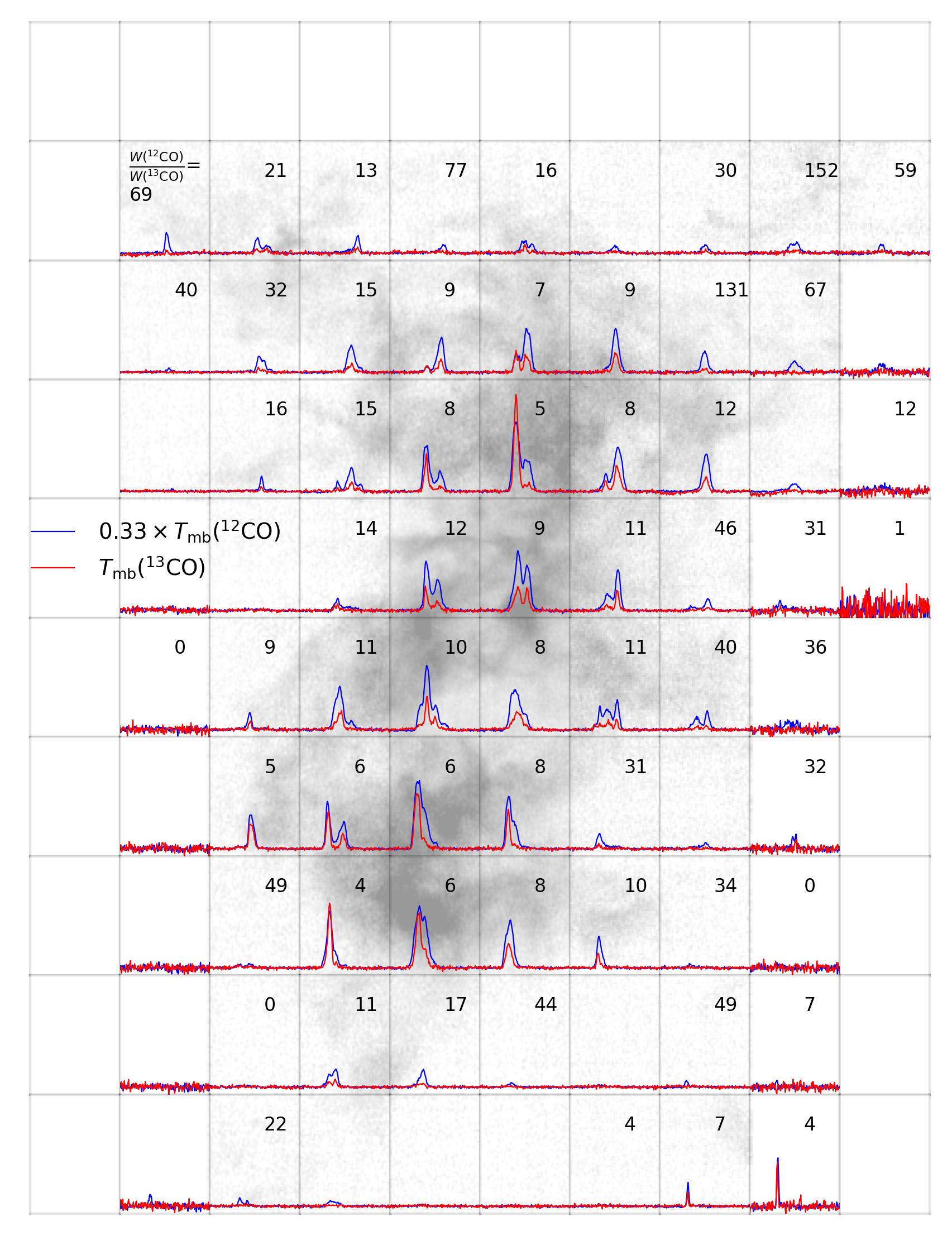}
    \caption{Spatial variation in $^{12}$CO and $^{13}$CO emission (blue and red respectively). The background is the $^{12}$CO integrated intensity. Each square shows the mean spectra within each grid position. The $W$($^{12}$CO)/$W$($^{13}$CO) line ratio for $>3 \sigma_{\rm rms}$ emission is shown in black. Note that $T_{\rm mb}$($^{12}$CO) has been multiplied by 0.33. Large line ratios $>50$ are seen when $^{12}$CO is detected but almost no $^{13}$CO.}
    \label{fig:lines_grid}
\end{figure*}

We plotted the average spectrum of both isotopologues for rectangular regions in MBM12 in Fig. \ref{fig:lines_grid}. The $^{12}$CO integrated intensity is shown in the background. The gridlines indicate the regions that were averaged. We also showed the mean $^{12}$CO/$^{13}$CO line ratio in each region.
\section{Molecular gas fraction}
\label{app:molecular_gas_fraction}
\begin{figure*}
    \centering
    \includegraphics[width=\linewidth]{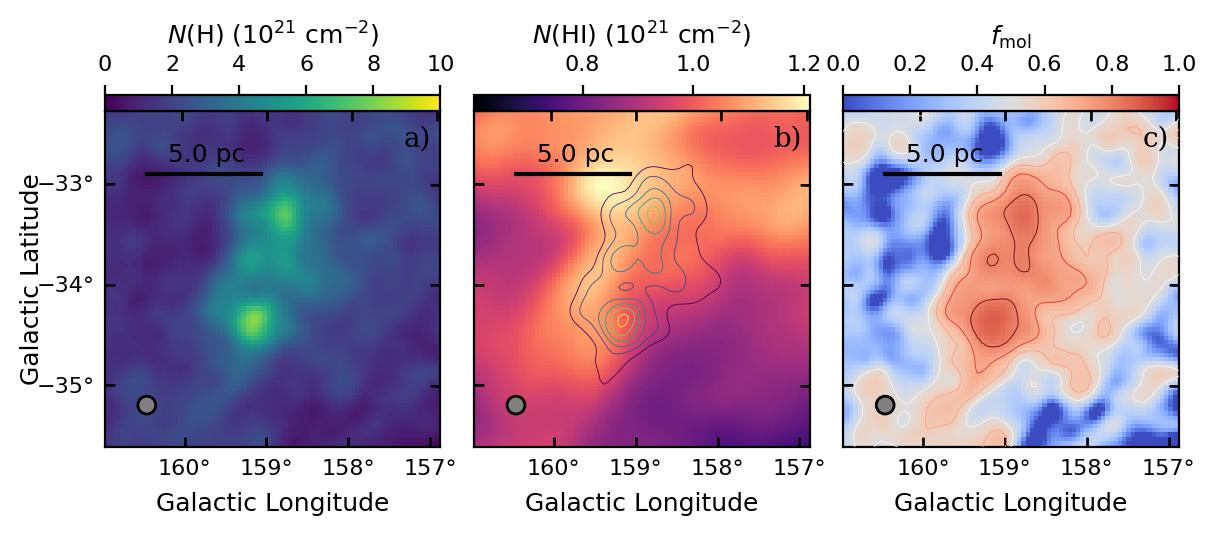}
    \caption{Estimate of molecular gas fraction in MBM12. Panel a shows total hydrogen column density from NICEST extinction maps. Panel b shows the atomic hydrogen column density from \citet{2016A&A...594A.116H}. Contours of $N$(H) $= [3, 4, 5, 6, 7,8] \times 10^{21}$ cm$^{-2}$ are shown in green. Panel c shows the molecular gas fraction from the combination of these observations with contours of $f_{\rm mol} = [0.5,0.6,0.7,0.8]$.}
    \label{fig:molecular_gas_fraction}
\end{figure*}

The total hydrogen content along the line of sight $N$(H) can be decomposed into ionized, atomic and molecular components:
\begin{equation}
    N({\rm H})  = N({\rm H}^+) + N({\rm HI}) + 2N({\rm H}_2)
\end{equation}
that implies the molecular fraction $f_{\rm mol}$ is given by:
\begin{equation}
    f_{\rm mol} = 1 - \frac{N({\rm HI}) + N({\rm H}^+)}{N({\rm H})}
\end{equation}
with the molecular hydrogen column density given by
\begin{equation}
    N({\rm H}_2) = \frac{f_{\rm mol}}{2} N({\rm H})
\end{equation}
We estimated the spatial variation in $f_{\rm mol}$ in MBM12 by combining total column density $N$(H) estimates from extinction with HI4PI all sky $N$(HI) estimates \citep{2016A&A...594A.116H}. We assumed $N$(H$^+$) to be negligible. Figure \ref{fig:molecular_gas_fraction} shows $N$(H), $N$(HI) and $f_{\rm mol}$ at $5'$ resolution. There is a shell of atomic hydrogen across the whole the east side of MBM12. The molecular gas fraction varies between 0.5 and 0.8 on subparsec scales.

\section{Dust opacity determination}
\label{app:dust_opacity_determination}
\begin{figure}
    \centering
    \includegraphics[width=0.8\linewidth]{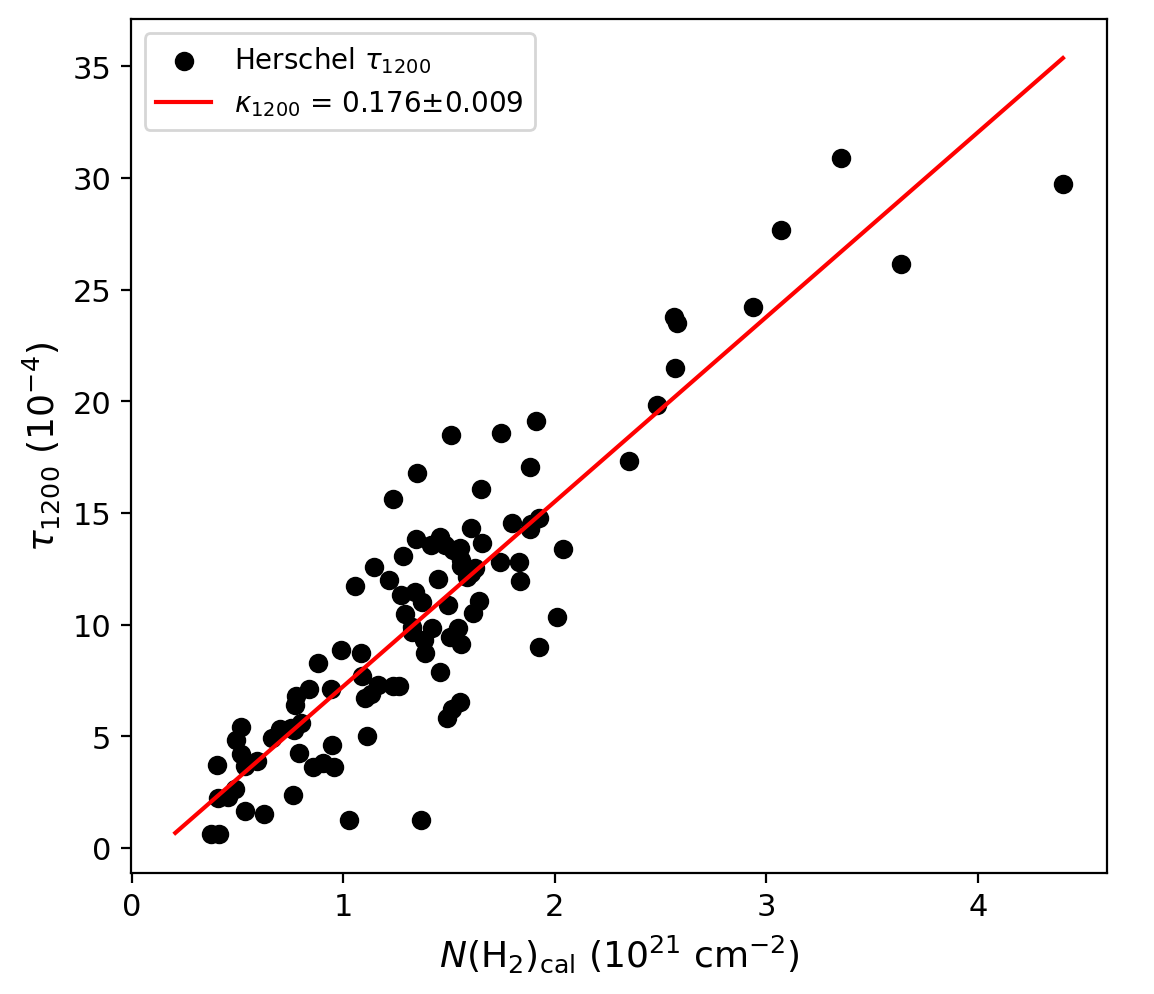}
    \caption{Dust opacity estimate at 1200 GHz for MBM12 using the column densities $N$(H$_2$)$_{\rm cal}$ of \citet{2017A&A...601A..78R} as a calibrator in the case of a molecular gas fraction $f_{\rm mol}$ of 1. The resulting dust opacity is shown.}
    \label{fig:remy_opacity_fit}
\end{figure}

\begin{figure}
    \centering
    \includegraphics[width=\linewidth]{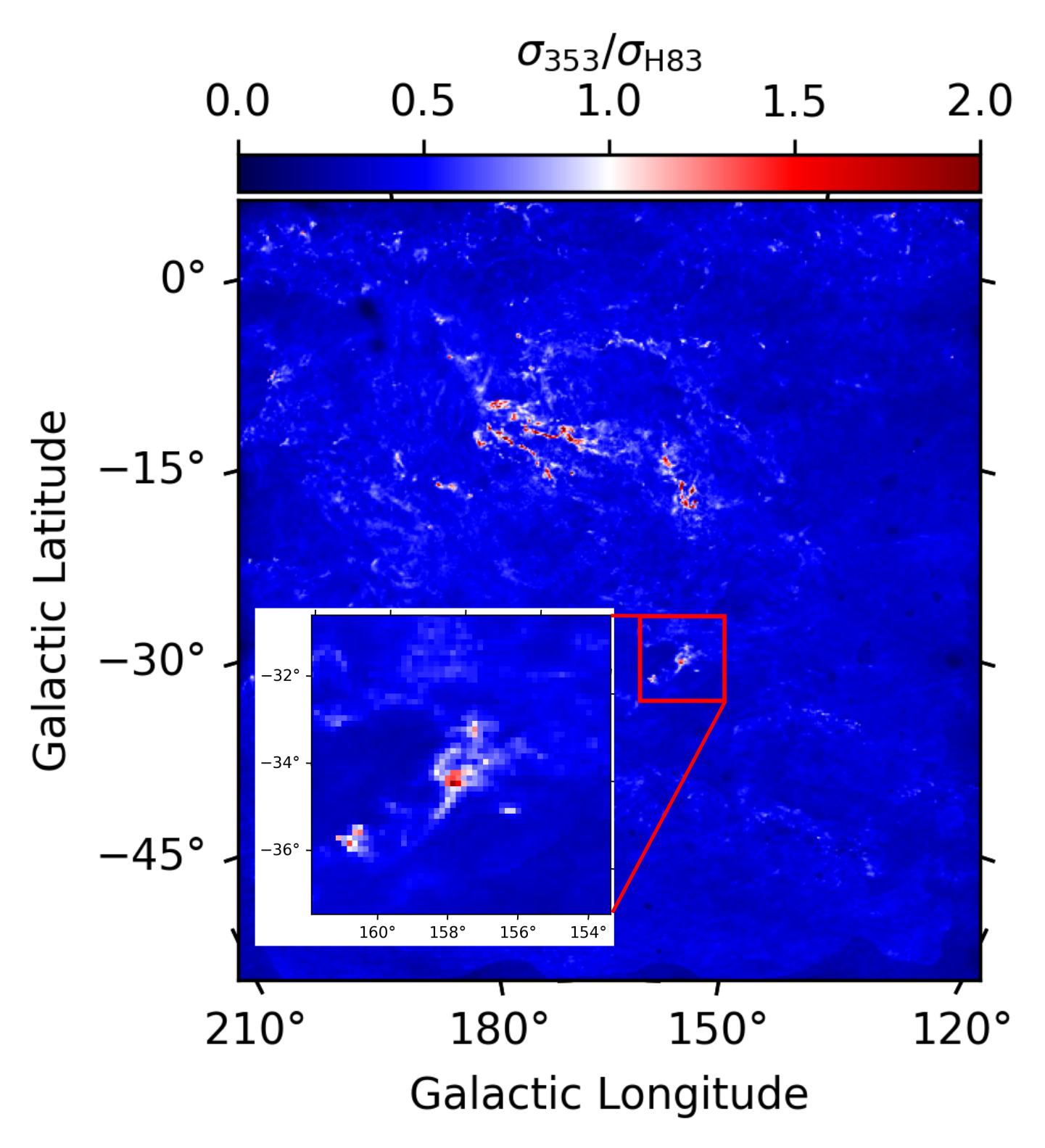}
    \caption{Dust opacity per nucleon, $\sigma_\nu$ from multi-wavelength modelling for anti-centre fields from \citet{2017A&A...601A..78R} as a ratio of the commonly used dust opacity from \citet{1983QJRAS..24..267H}, with $\beta = 2$. The insert is a zoom-in of MBM12, with the same colourscale.}
    \label{fig:remy_opacity_map}
\end{figure}

The dust opacity can be parameterized as:
\begin{equation}
    \label{eq:kappa}
    \kappa_\nu = \kappa_0\Bigg(\frac{\nu}{\nu_0}\Bigg)^{\beta},
\end{equation}
with $\nu_0$ a reference frequency, $\beta$ a power-law index, and $\kappa_0$ the dust opacity at the reference frequency. The values of $\kappa_0$ and $\beta$ are related to dust properties, such as composition, porosity, and size distribution.
In Table \ref{tab:dust_opacity_variation} we summarize ways that the dust opacity at 250 $\mu$m $\kappa_{\rm 1200}$ can be estimated. Estimates 1 and 2 consist of assuming a relevant literature value of $\kappa_0$ at $\nu_0$ with a power law exponent $\beta$ for Eq. \ref{eq:kappa}. The range of consequent values in $\kappa_{\rm 1200}$ originates from variation in $\beta$. The second class of $\kappa_{\rm 1200}$ estimates is to have a calibrator $N$(H) to calculate source-specific $\kappa_{\rm 1200}$. Estimate 3 is the average $\kappa_{\rm 1200}$ for the twelve molecular clouds in the survey of \citep{2022ApJ...931....9L}, calibrated with extinction. The uncertainties are the standard deviation within the twelve fields, while the range is determined by varying the $\beta$ of 353 GHz to 1200 GHz. Estimate 4 is the average and range of the surveyed values of $\tau(250$ $\mu$m)/$\tau(J)$ from \citet{2015A&A...584A..93J}, with the extinction to $N$(H) conversion used in Sect. \ref{subsec:nh2}, without accounting for $f_{\rm mol}$. Estimate 5 is also based on a calibrated $N$(H), but instead of extinction, is based on a linear model of multi-wavelength observations of anti-centre clouds, including MBM12 \citep[$\gamma$-rays, CO, FIR emission, HI,][]{2017A&A...601A..78R}. They constructed wide-field $40^\circ \times 40^\circ$ maps at $0.175^\circ$ resolution by linear modeling of HI, CO and free-free emission against a combination of {\it Planck} $\tau_{353}$ and Fermi-LAT $\gamma$-ray observations. One can calculate the dust opacity per nucleon $\sigma_\nu = \kappa_\nu\mu_{\rm H}$m$_{\rm p}$ from the \citet{2017A&A...601A..78R} $N$(H) map. In Fig. \ref{fig:remy_opacity_fit} we show an example of the estimate of $\kappa_{\rm 1200}$ by Eq. \ref{eq:tauNH2} with $N$(H$_2$)$_{\rm cal}$ in the case of $f_{\rm mol} = 1$. Fig. \ref{fig:remy_opacity_map} shows the spatial variations in the dust opacity per nucleon, $\sigma_{\rm 353}$, compared to the \citet{1983QJRAS..24..267H} for $\beta = 2$. The value is close to unity at the edges of MBM12, but with a change of a factor of two within the cloud. Estimate 6 is the values we estimated in Sect. \ref{subsec:nh2} of this work.

\section{Image segmentation with $X$(CO)}
\label{app:XCO-segmentation}
We fit the $X$(CO) PDFs with a lognormal distribution, given by \begin{equation}
    f_{\rm LN}(X({\rm CO})) = \frac{A_{\rm LN}}{X({\rm CO})\sigma_{\rm LN} \sqrt{2\pi}}{\rm exp}\Big( \frac{({\rm ln}[X({\rm CO})]-\mu_{\rm LN})^2}{2\sigma^2_{\rm LN}} \Big)
    \label{eq:lognormal}
\end{equation}
and with a truncated powerlaw
\begin{equation}
    f_{\rm PL} = A_{\rm PL} X({\rm CO})^{\alpha_{\rm PL}}
    \label{eq:powerlaw_XCO}
\end{equation}
where $X$(CO) was in units of $X_{\rm Gal}$.
\begin{table*}[]
    \centering
    \caption{Fitted parameters on the $X$(CO) PDFs in MBM12 (Fig. \ref{fig:herschel_xfactors} and Eqs. \ref{eq:lognormal} and \ref{eq:powerlaw_XCO}).}
    \begin{tabular}{llllllll}
    \hline
     ID & PDF Quantity & $A_{\rm LN}$ & $\mu_{\rm LN}$ & $\sigma_{\rm LN}$ & $A_{\rm PL}$&  $\alpha_{\rm PL}$ & PL Range \\
     & & (Counts) & (ln[$X_{\rm Gal}$]) & (ln[$X_{\rm Gal}$]) & (Counts) &  & ($X_{\rm Gal}$)  \\
    \hline
    1 & $X$($^{12}$CO) &  $1358\pm32$  & $-0.17 \pm 0.01$ & $0.32\pm0.01$ & $1907\pm35$ & $-1.21\pm0.03$ & [1.0, 10.0]  \\
    2 & $X$($^{13}$CO) & $3757\pm100$ & $1.71\pm0.01$ & $0.37\pm0.01$  & $2573\pm217$ & $-0.78\pm 0.04$ & [6.5, 20.0]  \\
    \hline
    \end{tabular}
    \tablefoot{The PL Range column shows the $X$(CO) ranges used for the truncated power law fit.}
    \label{tab:xco-fitted-parameters}
\end{table*}
\begin{figure*}
    \centering
    \includegraphics[width=\linewidth]{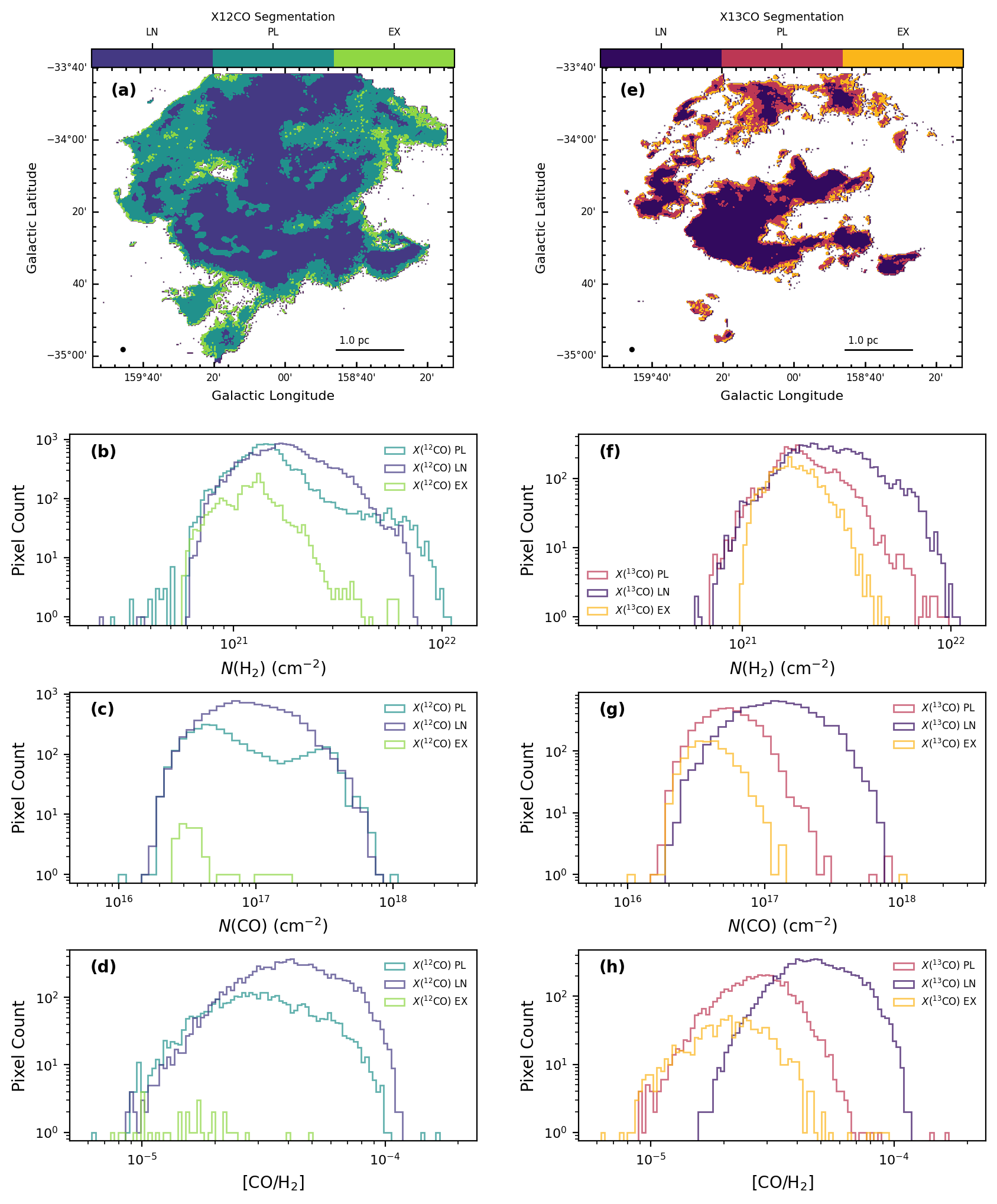}
    \caption{Single-isotopologue segmentation of MBM12 based on $X$(CO) PDF decomposition. Top panels: Spatial distribution of  lognormal (LN, left) and power-law (PL, right) classified regions for $^{12}$CO (panels a$-$d) and $^{13}$CO (panels e$-$h), overlaid on the $N$(H$_2$) map. The excess (EX) regions appear predominantly at cloud edges. Lower panels: Histograms of $N$(H$_2$), $N$(CO), and [CO/H$_2$] for pixels in each classification}
    \label{fig:xco_segmentation}
\end{figure*}
The fitted parameters for MBM12 are shown in Tab. \ref{tab:xco-fitted-parameters}. After this PDF decomposition, we investigated whether the statistical modes of the $X$(CO) PDF may correspond to different physical conditions. We generated binary masks according to whether the PDF is lognormal (LN), powerlaw (PL) or in the excess region (EX), according to these criteria
\[
\text{Mask (l,b)} = 
\begin{cases}
\text{LN} & \text{if } X({\rm CO}) \leq X_{\text{LNMax}} \\
\text{PL} & \text{if } X_{\text{LNMax}} < X({\rm CO}) \leq X_{\text{PLMax}} \\
\text{EX} & \text{if } X({\rm CO}) > X_{\text{PLMax}} 
\end{cases}
\]
where $X_{\rm LNMax} = X_{\rm LNMode} \cdot {\rm exp}(1.1775 \cdot \sigma_{\rm LN})$ where $X_{\rm LNMode} = {\rm exp}(\mu_{\rm LN} - \sigma^2_{\rm LN})$ is the distribution peak of a lognormal. The 1.1775 factor is half the FWHM of a lognormal in log-log space. The value of $X_{\rm PLMax}$ is the maximum of the PL Range column of Tab. \ref{tab:xco-fitted-parameters}, just before the excess peak (Fig. \ref{fig:herschel_xfactors}). Figures \ref{fig:xco_segmentation}\,a and \,e show the three masks for the two isotopologues. We also plotted the histograms of $N$(H$_2$), $N$(CO) and [CO/H$_2$] for these regions (panels b$-$d and f$-$h for $^{12}$CO and $^{13}$CO respectively). Regarding the maps we note that the EX mask is always at the edge of the cloud. For $^{13}$CO it follows the order of EX $\rightarrow$ PL $\rightarrow$ LN from outside to inside the cloud. However, for $^{12}$CO, there is a filamentary structure on the inside that is PL, which is not seen in $^{13}$CO. This is likely related to $^{12}$CO transitioning to high optical depths, while $^{13}$CO is still optically thin. The $N$(CO) PDFs are very different between masks. The LN mask in both isotopologue has a broad lognormal distribution centred around $10^{17}$ cm$^{-2}$, while the PL mask is bimodal for $^{12}$CO and single peaked for $^{13}$CO. The EX mask has the lowest peak value for both isotopologues. The abundance PDFs show mostly lognormal distributions, with a possible powerlaw at the low abundance end for $^{12}$CO, while for $^{13}$CO the three distributions are lognormal, with the peak abundance decreasing in the order LN $\rightarrow$ PL $\rightarrow$ EX. We consider these masks to segment the MBM12 map into distinct physical regions based on CO chemistry and optical depth effects. The detailed reasons why $X$(CO) change in these statistical behaviour requires additional investigation over multiple clouds and other spectral lines. Comparison with radiative transfer and chemistry simulations will also be helpful. 
\section{Considerations when interpreting $\alpha_{\rm vir}$}
\label{app:virial_analysis_interpretation}
Most forms of the observational $\alpha_{\rm vir}$ for molecular clouds are the ratio of internal kinetic energy to gravitational energy \citep[e.g.][]{1992ApJ...395..140B}. There has been an ever-evolving discussion on the (ir)relevance of virial parameters. One's interpretation of $\alpha_{\rm vir}$ varies based on whether one assumes static or dynamic MCs. Forty years ago, it was common to conclude that molecular clouds are in virial equilibrium from the Larson $\sigma_{\rm v} \propto R^{0.5}$ scaling relation, and to use that to calculate the cloud virial mass \citep{1987ApJ...319..730S,1988ApJ...334L..51M,1989ApJ...337..704L}. From similar static assumptions, some have interpreted $\alpha_{\rm vir} < 1$ as a sign of other supporting forces such as internal turbulence or magnetic fields \citep{1992ApJ...395..140B,2013ApJ...779..185K}. Those discussed above used an assumption of virial equilibrium to either calculate masses or infer some unmeasured force. For the dynamic assumption of MCs, the value of $\alpha_{\rm vir}$ can be used as a measure of (un)boundedness of a cloud. There the assumption is that internal kinetic and gravitational energies are the only relevant energies, and so $\alpha_{\rm vir} > 1-2$ can be used as a simple binary measure for gravitational boundedness. Many users of this approach do flag its tentative nature \citep[][]{2008ApJ...679.1338R,2016ApJ...833..204F,2024ApJ...969...70F,2025ApJ...993..193O}. Strictly speaking, this approach in particular has no way to distinguish if the cloud is gravitationally collapsing (no supporting forces), in pressure equilibrium (unobserved supporting forces), dispersing (unobserved forces stronger than gravity) or fragmenting \citep{1987A&A...171..225C}. Virial parameters are a local measure and ignore external pressure, large-scale velocity gradients, and rotation \citep{2006MNRAS.372..443B,2023ASPC..534....1C}. There are also observational biasing effects, such as background subtraction, sampling due to line excitation and variable emission-to-mass conversion factors that may lead to artificially small $\alpha_{\rm vir}$ \citep{2021ApJ...922...87S}. Some have therefore proposed alternative observational estimators of gravitational stability \citep{2015A&A...578A..97L,2025OJAp....8E..91K}. The observational virial parameter from PPV cubes is different from a 3D virial parameter in a simulation due to projection. In simulations, one can directly calculate gravitational, kinetic and magnetic energies. Some simulations of MCs suggest that observational virial parameters are not associated with cloud stability at all, due to tidal and projection effects \citep{2007ApJ...661..262D,2021ApJ...911..128K,2022MNRAS.517..885O,2025ApJ...988..266L}. These considerations show that one should be upfront about assumptions when doing virial analysis, and that the conclusions from a virial analysis alone are only tentative without combination with independent measures. 
\end{appendix}
\end{document}